\documentclass[twocolumn]{aastex63}

\usepackage{amsmath}
\usepackage{enumerate}
\usepackage{rotating}

\newcommand{\appropto}{\mathrel{\vcenter{
  \offinterlineskip\halign{\hfil$##$\cr
    \propto\cr\noalign{\kern2pt}\sim\cr\noalign{\kern-2pt}}}}}

\received{November 5, 1955}
\revised{October 25, 1985A}
\accepted{October 21, 2015}
\submitjournal{ApJ}

\shorttitle{Sample article}
\shortauthors{Martin \& Fabrycky}
\graphicspath{{./}{figures/}}

\begin{document}

\title{}

\title{Searching for Small Circumbinary Planets \\
I. The STANLEY Automated Algorithm and No New Planets in Existing Systems}
\correspondingauthor{David V. Martin}
\email{martin.4096@osu.edu}

\author[0000-0002-7595-6360]{David V. Martin}
\affiliation{Department of Astronomy, The Ohio State University \\
4055 McPherson Laboratory, Columbus, OH 43210, USA}
\affiliation{Fellow of the Swiss National Science Foundation} 
\affiliation{Department of Astronomy and Astrophysics, University of Chicago \\
5640 S Ellis Ave, Chicago, IL 60637, USA}

\author[0000-0003-3750-0183]{Daniel C. Fabrycky}
\affiliation{Department of Astronomy and Astrophysics, University of Chicago \\
5640 S Ellis Ave, Chicago, IL 60637, USA}



\begin{abstract}

No circumbinary planets have been discovered smaller than $3R_{\oplus}$, yet planets of this small size comprise over 75\% of the discoveries around single stars. The observations do not prove the non-existence of small circumbinary planets, but rather they are  much harder to find than around single stars, because their transit timing variations are much larger than the transit durations. We present \textsc{Stanley}: an automated algorithm to find small circumbinary planets. It employs custom methods to detrend eclipsing binary light curves and stack shallow transits of variable duration and interval using N-body integrations. Applied to the Kepler circumbinaries, we recover all  known planets, including the three planets of Kepler-47, and constrain the absence of additional planets of similar or smaller size. We also show that we could have detected $<3R_{\oplus}$ planets in half of the known systems. Our work will ultimately be applied to a broad sample of eclipsing binaries to (hopefully) produce new discoveries, and derive a circumbinary size distribution which can be compared to that for single stars.

\end{abstract}



\section{Introduction} \label{sec:intro}



\begin{figure*}  
\begin{center}  
\includegraphics[width=0.99\textwidth]{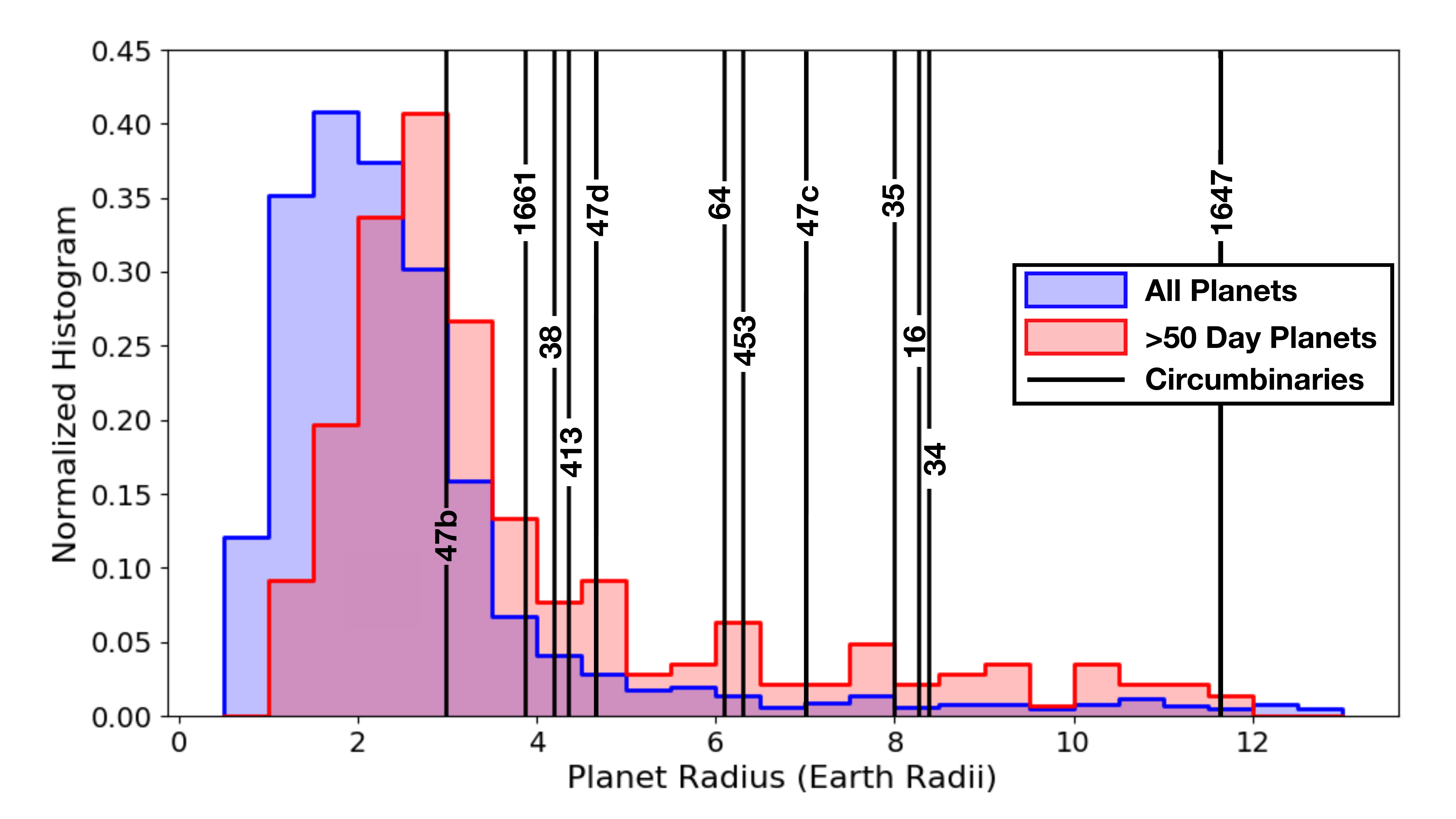}  
\caption{Histogram of confirmed Kepler objects under $13R_{\oplus}$ for all orbital periods (blue) and $P_{\rm p}>50$ days (red), where the latter corresponds to the period range of discovered circumbinary planets. Black vertical lines denote the published Kepler circumbinary planet radii, with numbers denoting their Kepler number. The circumbinary planet radii are relatively large compared with planets around single stars, both with and without the 50-day period cut. Note that the typically larger radii of the $>50$ day sample is largely an observational bias due to longer period planets producing less transits. Nevertheless, roughly half of the $>50$ day planets around single stars are smaller than all of the circumbinary planets. Data taken from the NASA Exoplanet Database  on May 4 2020, using all Kepler Objects of Interest (KOIs) with ``CONFIRMED'' status.}
\label{fig:single_star_abundance}
\end{center}  
\end{figure*}

All transiting circumbinary planets discovered to date are larger than $3R_{\oplus}$ \citep{martin2018,welsh2018}. This is in contrast to planets around single stars, for which smaller planets are significantly more abundant than their gas giant counterparts (Fig.~\ref{fig:single_star_abundance}, \citealt{petigura2013,fulton2017}). On the surface this looks like a stark difference in populations, potentially providing an insight into planet formation in different environments. However, detection limitations make such a comparison is premature.

Around single stars, planetary transits occur on regular intervals. This regularity is the basis of common detection methods such as Boxed Least Squares (BLS, \citealt{kovacs2002}), where the light curve is phase-folded on a periodic interval to check for significance within a folded ``box". Phase-folding transits becomes essential for small planets, as their shallow transit depths might be individually insignificant, but detectable as an ensemble.

Deviations from strict periodicity are called Transit Timing Variations (TTVs). For single stars these are typically on the order of seconds or minutes \citep{holman2005,agol2005}, but for circumbinary planets they can be on the order of {\it days}  or even {\it weeks} \citep{armstrong2013}. The orbital motion of the inner binary is the primary component of these large TTVs, which are on order ${\rm TTV}\approx (P_{\rm bin}^2P_{\rm p})^{1/3}/(2\pi)$. This relative motion also induces significant transit duration variations (TDVs); a given planet may have a transit of a few hours followed by a transit of over a day in length \citep{schneider1990,kostov2013}. There are secondary, but significant, contributions to the TTVs and TDVs from three-body interactions, which occur on both long secular timescales (e.g. nodal precession and apsidal advance) and shorter orbital timescales \citep{schneider1994,mardling2013,martin2014,kostov2014}.

Since the TTVs are typically much larger than the transit durations,  phase-folding photometric data on a fixed period tends to wash away circumbinary transit signals rather than stacking them coherently. This invalidates automated techniques used on single stars. 

As of now, the most popular and successful means of finding circumbinary planets  has been to search for their transits by eye\footnote{Although see the semi-automated method first presented in \citet{kostov2013} in Sect.~\ref{subsubsec:discussion_comparison_kostov}.}. This has accounted for all of the Kepler discoveries (starting with Kepler-16, \citealt{doyle2011}) and the only TESS discovery (TOI-1338/EBLM J0608-59, \citealt{kostov2020}). There are two main limitations with this method. First, it requires the individual transits to be discernable by eye, inhibiting the discovery of small planets. Second, it is difficult to quantify the efficiency of this method and constrain non-detections.


There have been efforts to  develop automated detection algorithms specific to circumbinary planets: \citet{jenkins1996}, \citet{ofir2008} (CBP-BLS), \citet{carter2013} (QATS), \citet{armstrong2014}, \citet{klagyivik2017} and \citet{windemuth2019} (QATS-EB). Some of these algorithms have demonstrated a potential sensitivity to small planets but have not yet been applied to a large sample \citep{jenkins1996,windemuth2019}. On the other hand, there have been comprehensive applications to Kepler \citep{armstrong2014} and CoRoT \citep{klagyivik2017}, but only with  sensitivity to gas giants. There have also been advances in modelling the dynamics of circumbinary planets \citep{leung2013,mardling2013,georgakarakos2015}, but they are yet to be applied in the detection of new planets. Finally, every published circumbinary planet discovery has included a photodynamical model fitted to the data, which fully encapsulates the three-body geometry and dynamics. However, such models are only to characterise  and confirm the planet {\it after} the planet has been detected by eye, meaning that the final fit can be made over a narrow, computationally-tractable parameter space.

We present a new solution to the problem of automatically {\it finding} circumbinary planets. The \textsc{Stanley} algorithm invokes a brute-force grid search with an N-body integrator. This most closely resembles the oldest of the above works - \citet{jenkins1996} - whereas subsequent work developed clever approximations to the transit timing to speed up processing. The advantage of using an N-body  code is that all dynamical and geometric effects on the transit signature are naturally accounted for. The trade-off is unsurprisingly a decrease in processing speed, and so we develop an optimized search grid so that the method remains tractable. \textsc{Stanley} also incorperates bespoke detrending methods specific to eclipsing binaries, to best illuminate shallow transits of variable duration in the presence of stellar and instrumental variability of typically greater magnitude.

In this paper we demonstrate the new \textsc{Stanley} algorithm and apply it to the known Kepler circumbinary planets. We both recover the true planets and to set detection limits for each system by recovering simulated planets with variable transit depths. We also search for additional planets in each system. This paper is a prelude to a second paper, in preparation, where we will apply \textsc{Stanley}  to a broader sample of eclipsing binaries to constrain the size distribution of circumbinary planets, potentially including new discoveries.

This paper is structured as follows. We discuss the treatment and detrending of the lightcurves in Sect.~\ref{sec:detrending}. The search algorithm is presented in Sect.~\ref{sec:search}. We then apply this algorithm to the detection of known Kepler planets in Sect.~\ref{sec:results} to find the known planets, search for additional planets to place detection limits. Finally in Sect.~\ref{sec:discussion} we discuss applications of our work, future improvements, and provide a more in depth comparison of our method to existing circumbinary planet techniques, before concluding in Sect.~\ref{sec:conclusion}.

\section{Detrending and data processing}\label{sec:detrending}

\subsection{SAP Flux}\label{subsec:detrending_sapflux}

The starting point is the SAP (Simple Aperture Photometry) flux downloaded using the \textsc{Lightkurve} package \citep{lightcurve2018}\footnote{\url{https://github.com/KeplerGO/lightkurve} \textsc{Lightkurve} makes use of several packages: \textsc{astropy} \citep{astropy2013}, \textsc{astroquery} \citep{ginsberg2019} and \textsc{celerite} \citep{foreman-mackey2017}.}. \textsc{Lightkurve} has four settings for discriminating data based on the data quality flags: ``none'', ``normal'', ``hard'' and ``hardest''. We conservatively used the ``hard'' setting because from inspection of some of the light curves the default ``normal'' setting included a noticeable number of outlier points. The ``hardest'' setting is not recommended  according to the \textsc{Lightkurve} documentation.

We only downloaded long-cadence (29.4 minutes timesteps) data for all available Kepler quarters. No short-cadence data were used because whilst our search and detrending algorithms are not restricted to any specific observing cadence, they are not adapted to a variable cadence. The data were stitched together using default \textsc{Lightkurve} tools.

Like what was done for the successful by-eye searches for circumbinary planets, we avoid using the PDC (Pre-search Data Conditioning)\footnote{Jerry Orosz and Bill Welsh, private communication.} and do our own de-trending.

\subsection{Orbital and physical parameters}\label{subsec:detrending_parameters}

We obtain orbital parameters from the Villanova eclipsing binary catalogue \citep{prsa2011,kirk2016}\footnote{\url{http://keplerebs.villanova.edu/}}, which contains almost 3,000 entries. The data needed are the period ($P_{\rm bin}$), the  primary eclipse time ($T_{\rm 0,pri}$), the primary and secondary  scaled eclipse durations\footnote{i.e. scaled to the binary period.} ($w_{\rm pri}$ and $w_{\rm sec}$) and the  scaled separation between primary and secondary eclipses:

\begin{equation}
    \label{eq:eclipse_separation}
    s=\frac{T_{\rm 0,sec} - T_{\rm 0, pri}}{P_{\rm bin}}.
\end{equation}
 For a circular orbit ${\rm s}=0.5$. From this we calculate $e_{\rm bin}\cos\omega_{\rm bin}$ from the phase offset of eclipses,

\begin{equation}
    \label{eq:ecosw}
    e_{\rm bin}\cos\omega_{\rm bin} = \frac{\pi}{2}\left(s-\frac{1}{2}\right),
\end{equation}
and $e_{\rm bin}\sin \omega_{\rm bin}$ from the relative eclipse widths,

\begin{equation}
    \label{eq:esinw}
    e_{\rm bin}\sin \omega_{\rm bin} = \frac{w_{\rm sec} - w_{\rm pri}}{w_{\rm sec}+w_{\rm pri}}.
\end{equation}
Typically $e_{\rm bin}\cos\omega_{\rm bin}$ is better constrained than $e_{\rm bin}\sin\omega_{\rm bin}$. We then calculate $e_{\rm bin}$ and $\omega_{\rm bin}$ from these two relations.  Note that Eqs.~\ref{eq:ecosw} and \ref{eq:esinw} are approximations that assume the binary is inclined at exactly $90^{\circ}$. 

We take the masses and radii for the primary and secondary stars ($m_{\rm A}$, $m_{\rm B}$, $R_{\rm A}$ and $R_{\rm B}$ from the catalog produced by \citet{windemuth2019}, which contains data for over 700 of the Kepler eclipsing binaries. The Windemuth catalog is based on the Kepler photometry and GAIA stellar characterisation. They make calibrations with the binaries with masses calculated in the traditional way as a double-lined spectroscopic binary (e.g. \citealt{matson2017}) and show an accuracy to $\sim 10-20\%$. 

Note that we use the Windemuth  catalog values even when running searches on the known Kepler circumbinary planets. The discovery papers for those planets undoubtedly contain more precise stellar values, as they are based on radial velocities, eclipse timing variations and circumbinary planet transits, none of which are accounted for by \citet{windemuth2019}. However, we use the Windemuth catalogue in this paper in anticipation of our second paper, where we blindly search for new planets, most of which around binaries for which the \citet{windemuth2019} parameters are the best known. We will demonstrate that even with less precise stellar parameters we recover the known planets with high significance.

\subsection{Removing stellar eclipses}\label{subsec:detrending_stellar_eclipses}

The fraction of time the binary spends in eclipse (primary or secondary) is $\sim(R_{\rm A}+R_{\rm B})/a_{\rm bin}$ for circular orbits and $I_{\rm bin}=90^{\circ}$. This is significant, particularly for short-period binaries. Ideally, one could model the stellar eclipses, subtract this model, and be left with a light curve without large chunks removed. If done properly, blended planetary transits could be discovered. In practice though, the eclipses are orders of magnitude deeper than the planetary transits we are seeking. Any slight imperfections in the binary fit might be tiny with respect to the eclipse but could be comparable to planetary transits, and in our experience attempts to model the eclipses resulted in artifact ``transits''. We hence avoided this, and simply removed the in-eclipse data. 

 In each light curve we isolate the observing cadences in eclipse according to the Villanova catalog values of the primary and secondary eclipse times ($T_{\rm 0,pri}$ and $T_{\rm 0,sec}$, calculated from $s$ in Eq.~\ref{eq:eclipse_separation}) and durations ($w_{\rm pri}$ and $w_{\rm sec}$). We then remove these cadences. The Villanova catalog values are fitted allowing for eccentric binaries and hence potentially substantially different primary and secondary eclipse durations.

\subsection{Transit duration variations -- theory}\label{subsec:detrending_transit_duration_theory}

We approximate the transit duration by

\begin{equation}
    \label{eq:tau}
    \tau = \frac{2\sqrt{\left(R_{\rm A,B}+R_{\rm p}\right) - \left(bR_{\star}\right)^2}}{v_{\rm p} - v_{\rm A,B}},
\end{equation}
where $b$ is the impact parameter, $R_{\rm A,B}$ is the radius of the star being transited and $v_{\rm p}$ and $v_{\rm A,B}$ are the planet and star transverse velocities across the sky, respectively, calculated by

\begin{equation}
\begin{split}
\label{eq:velocity}
    v_{\rm A,B} & = \frac{m_{\rm B,A}}{m_{\rm A} + m_{\rm B}}\frac{2\pi a_{\rm bin}}{P_{\rm bin}}\sin\theta_{\rm A,B},\\
    v_{\rm p} & = \frac{2\pi a_{\rm p}}{P_{\rm p}}\sin\theta_{\rm p}.
\end{split}
\end{equation}
 These calculations assume perfectly edge-on binary and planet orbits with respect to the observer. We discuss this assumption in Sect.~\ref{subsubsec:search_grid_ignore}.

\begin{figure}  
\begin{center}  
\includegraphics[width=0.49\textwidth]{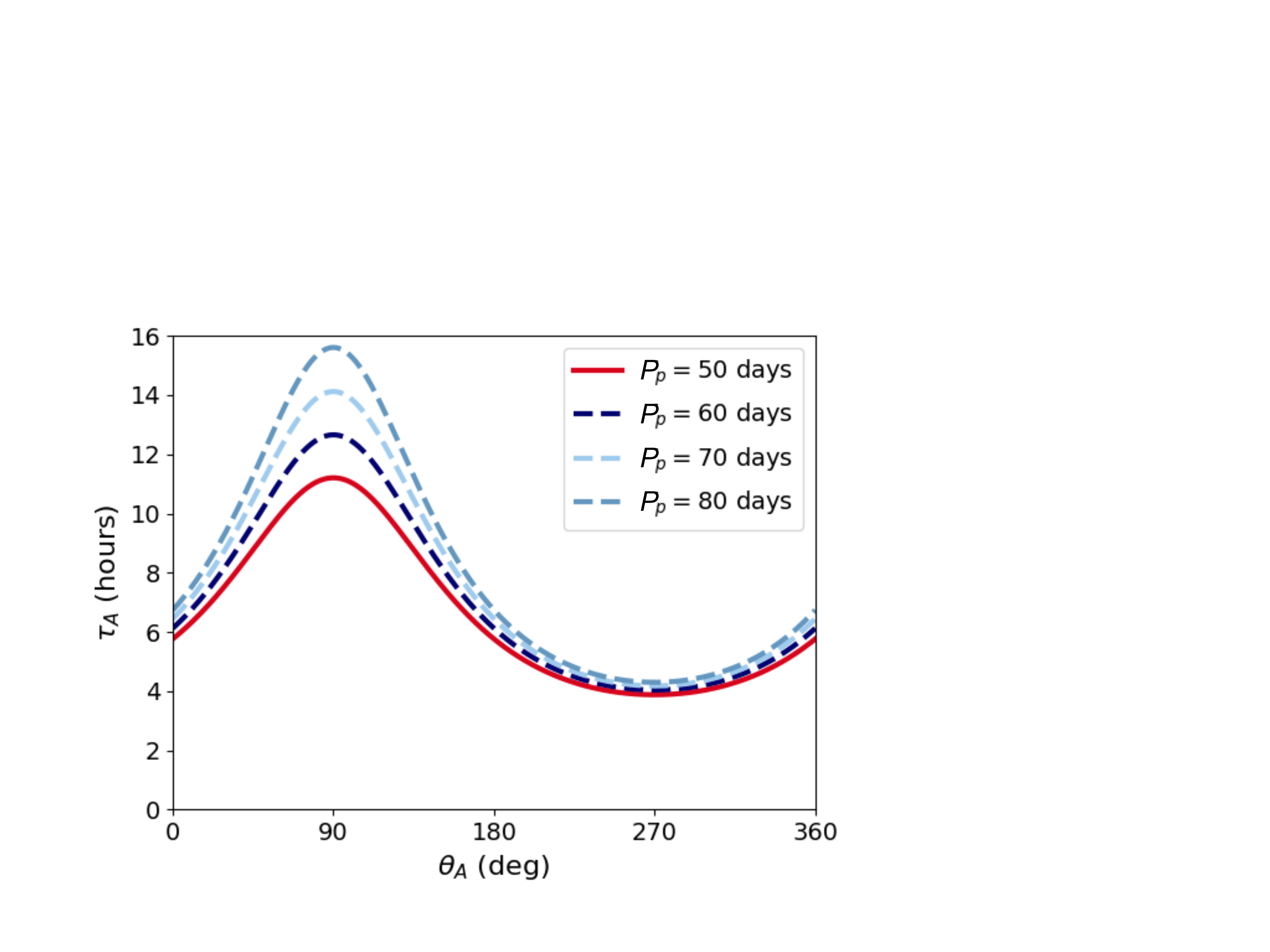}  
\caption{Circumbinary planet transit duration on the primary star for  Kepler-47b (red solid curve) and longer planet periods (blue dashed curves). This is calculated according to the simplified Eq.~\ref{eq:tau} assuming $\theta_{\rm p}=\pi/2$ at transit, $b=0$ and both orbits are circular. This variation of $\tau$ with the binary orbital phase is the dominant contribution to the TDVs.}
\label{fig:transit_duration}
\end{center}  
\end{figure}

 Around single stars there may be small changes to $\tau$ (TDVs) in the case of multiple planets or precession due to tides or general relativity, but these are typically on the order of seconds or minutes. Around binary stars  $\tau$ may vary considerably, owing to {\it four} effects. First, as seen in Eq.~\ref{eq:velocity} the velocity of the star $v_{\rm A,B}$ is a function of the binary phase, which could be anywhere between $0$ and $2\pi$ when the planet transits. In Fig.~\ref{fig:transit_duration} we show the variation of $\tau_{\rm A}$ as a function of $\theta_{\rm A}$.

Second, the phase of the planet $\theta_{\rm p}$ can no longer be assumed to be very close to $\pi/2$ at transit. Unlike single stars, for which the planetary velocity is therefore the same at each transit according to Eq.~\ref{eq:velocity}, for circumbinaries there is range of possible velocities too (albeit a relatively smaller range than for the star). 

The first two effects are applicable to a non-interacting pair of static Keplerians. However, in circumbinary systems the gravitational potential of the binary perturbs the planetary orbit on both short, orbital timescales and on longer secular timescales. This produces a third element to the TDVs. These variations are amplified in the presence of eccentric binary and/or planet orbits.

 For the fourth and final effect, we note that all of the above is applicable to planets on coplanar orbits with respect to the binary. If the planet is misaligned, even by a few 10th's of a degree, then an additional source of TDVs will arise primarily due to variations in the impact parameter, $b$, which may be so large that the planet actually stops transiting for long periods of time,  as explored theoretically in \citealt{schneider1990,martin2014,martin2017} and observed in Kepler-47 \citep{orosz2012,orosz2019}, Kepler-413 \citep{kostov2014} and Kepler-1661 \citep{socia2020}.

In circumbinary transit searches these large TDVs can be beneficial. In concert with the TTVs, they form a  ``smoking gun'' signature which is very hard to mimic by false positives (estimates for the false positive rate given in \citealt{kostov2020}). When it comes to detrending the data, however, these TDVs present a challenge.

\subsection{Transit duration variations -- in the context of detrending}\label{subsec:detrending_transit_duration_detrending}

 Every light curve contains a multitude of  trends, both instrumental and astrophysical, in addition to the transits we are trying to identify. The task at hand is to remove all variations except transits,  such that these transits are better illuminated. This process of detrending is the application of a high-pass filter to remove the low-frequency signals caused by instrumental and stellar variations, whilst preserving the relatively sharp, short-timescale transit features. Detrending algorithms, whether it be the commonly-used Savitsky-Golay filter \citep{savitzky1964}, simpler mean and median filters, or the cosine \citep{mazeh2010} and Tukey's biweight \citep{mosteller1977} filters which we will be using, all act according to a ``window length''. In brief, the algorithms are designed to remove signals on a timescale longer than some multiple of the window length, and preserve those on a shorter timescale. 

For a planet of a given orbital period transiting a single star its constant duration can be calculated, and then this would dictate the detrending window length such that transits are preserved. For a circumbinary planet, even at a fixed orbital period, the durations vary by almost an order of magnitude (e.g. Fig.~\ref{fig:transit_duration}).

A detrending window adapted for the short circumbinary transits risks removing the longer duration ones, which are the transits which would carry the most signal-to-noise weight when it comes to detecting the planet. On the other hand, if the detrending window is widened to account for transits of all durations, then we risk a light curve full of remaining, unwanted variations. 

Our method in \textsc{Stanley} is to take each time in the light curve and calculate how long a hypothetical planet transit would be if it occurred then. This allows us to vary the detrending window accordingly.

The binary and stellar parameters are known from the Villanova and Windemuth catalogues. To predict the parameters of the as of now undetected planet we use $P_{\rm p} = 6.1P_{\rm bin}$, which roughly corresponds to most of the known Kepler circumbinary planets \citep{martin2018} and is optimized for the detection of the shortest possible period planets\footnote{Planets closer than $\sim 4P_{\rm bin}$ develop an unstable orbit on short timescales \citep{dvorak1984,holman1999,mardling2001,quarles2018}.}, which are also going to be the most detectable since they will have the highest number of transits. As demonstrated in Fig.~\ref{fig:transit_duration}, $\tau$ is not a sharp function of $P_{\rm p}$, but we accept that some of the longer transits of longer-period planets may be  negatively impacted by the detrending.

The planet's starting phase is unknown, so we arbitrarily start it at $\theta_{\rm p}=0$. We set both the binary and planet inclinations to strictly be edge-on ($I_{\rm bin}=I_{\rm p}=\pi/2$). All of the known planets have a mutual inclination ($\Delta I$) less than 4$^{\circ}$ so the assumption of coplanarity is reasonable. However, even small misalignments will create transit impact parameters which are both non-zero and variable, and so assuming $\Delta I=0$ guarantees that we overestimate the transit duration by at least a small amount (see Sect.~\ref{subsubsec:search_grid_ignore}). The binary eccentricity and apsidal alignment use the values from the Villanova catalogue, whereas the planet is set to circular\footnote{We only assume a circular planet for the sake of detrending, and not in the planet detection aspect of \textsc{Stanley} where we search over a grid of planet eccentricities. The eccentricity of a planet will affect its transit duration, but this is a smaller effect than that of the binary phase. We also note that the eccentricities of the known circumbinary planets are all below $<0.15$.} 

We calculate $\tau(t)$ with the N-body package {\sc Rebound} \citep{rein2012}\footnote{\url{https://github.com/hannorein/rebound}}, using its IAS15 integrator \citep{rein2015} to measure the star and planet velocities at each time step, which are converted to $\tau(t)$ using Eq.~\ref{eq:tau}.

\subsection{Cosine detrending}\label{subsec:detrending_cosine}

All of our light curve detrending is done using \textsc{WOTAN} \footnote{\url{https://github.com/hippke/wotan}}, which is a simple to use python package containing about two dozen detrending algorithms, and a companion paper comparing the methods' efficacy \citep{hippke2019b}. We use two of the \textsc{Wotan} algorithms in this paper. The first is an iterative cosine filter, which was originally developed by \citet{mazeh2010}. A series of sines and cosines is iteratively fitted and subtracted from the data until a convergence is found in the fit. 

Our first detrending step uses sinusoids because a lot of the stellar variation seen in close binary light curves (ellipsoidal variation, Doppler beaming/boosting and reflection \citealt{faigler2011}), as well as typical star spot-induced rotation modulations, is inherently sinusoidal.

\citet{hippke2019b} note that the iterative cosine method is similar to the CoFiAM (Cosine Filtering with Autocorrelation Minimization) algorithm developed by \citet{kipping2013} and then reproduced and published by \citet{rodenbeck2018}. CoFiAM was developed for the search of exomoons, which like our target small circumbinary planets exhibit shallow transits with a complex transit timing signature (see also \citealt{martin2017b}). We favour the \citet{mazeh2010} iterative cosine algorithm though since \citet{hippke2019b} demonstrated it was the most robust (as a function of exoplanet transits found) of the filters based on fitting splines, polynomials and sinusoids.

Our method is as follows:

\begin{enumerate}
    \item Using the variable transit duration $\tau(t)$ calculated in Sect.~\ref{subsec:detrending_transit_duration_detrending}, calculate the maximum transit duration, $\tau_{\rm max}$.
    \item Detrend the data using the cosine filter with a window length of $3\times\tau_{\rm max}$. All other parameters in the detrending are set to the \textsc{Wotan} defaults.
    \item Compute a Lomb-Scargle periodogram \citep{lomb1976,scargle82} of the detrended light curve.
    \item Compute  the periodogram power corresponding to a False Alarm Probability (FAP) of 1\%. This requires roughly white noise to be valid, which is the case here after the detrending.
    \item Check if there is any periodogram power above a 1\% FAP at a period longer than $\tau_{\rm max}$. If so, return to Step 2, but use now a window length of $2.5\times\tau_{\rm max}$. Repeat the subsequent steps,  subtracting from the the window length multiplier by 0.5 each time until the periodogram loses all power above a 1\% FAP for periods longer than $\tau_{\rm max}$. Note that whenever we need to decrease the window length we return to the original light curve (with eclipses removed), rather than cumulatively detrending the light curve with ever-decreasing window lengths. This is such that we are always detrending an inherently sinusoidal signal.
    \item If the periodogram still has power above a 1\% FAP  after the window length had decreased from 3 to $1\times \tau_{\rm max}$, return to Step 1 but instead of calculating $\tau_{\rm max}$, calculate $\tau_{\rm 75\%}$.  This value is defined as the transit duration for which 75\% of all possible binary phases at transit should produce a shorter transit.
    \item If the periodogram still has power above a 1\% FAP by the time the window length reaches $1\times \tau_{\rm 75\%}$ then we stop the cosine detrending, lest we risk filtering out a significant number of transits.
    
\end{enumerate}

\begin{figure*}  
\begin{center}  
\includegraphics[width=0.99\textwidth]{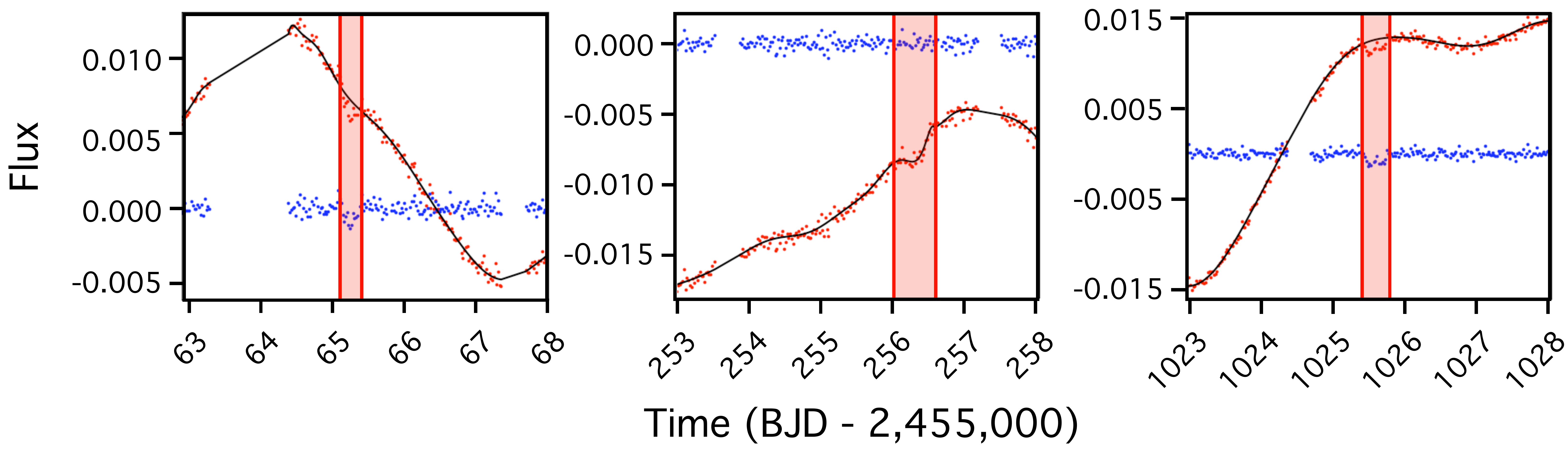}  
\caption{Example detrending of the Kepler-47 light curve with three (of over 20) transits. The raw light curve (after eclipses have been removed) is shown in red, and the detrended light curve is in blue, always at a normalized flux of 0. The black curve is the cosine filter trend. The transit is highlighted by the red shading, with a variable duration (a baseline of 5 days is shown in all three plots). In all three cases the stellar activity is over an order of magnitude greater than the transit depths. The first and third transit are preserved by the cosine detrending, whereas the second transit is significantly longer and gets removed by the cosine detrending. This transit removal is an unfortunate consequence of aggressive detrending, yet since the large majority of the Kepler-47 transits are untouched we will still be able to recover the planet (see  Fig.~\ref{fig:kepler_47_transit_map}).}
\label{fig:detrending_example}
\end{center}  
\end{figure*}

In Fig.~\ref{fig:detrending_example} we see the cosine filter being applied to Kepler-47. Most of the transits are preserved, but here we show here a set of three which include one long transit that gets removed by the filter.  This illustrates the main challenge with detrending: filtering away nuissance signals whilst preserving the planet transits. This challenge is exacerbated by the variable and often long duration circumbinary transits. We will see later though in Fig.~\ref{fig:kepler_47_transit_map} that of the over two dozen transits of Kepler-47b almost all of them are preserved, and in  Sect.~\ref{subsec:results_known} that we strongly detect the signal of the planet.

\subsection{Variable-window biweight detrending}\label{subsec:detrending_variable}

In our second stage of detrending we apply a Tukey's biweight filter \citep{mosteller1977} to the cosine-detrended light curve. This was found to be the most effective sliding filter by \citet{hippke2019b}, and significantly better than the much more commonly used Savitzky-Golay filter \citep{savitzky1964}. By using a sliding filter we should be more sensitive to trends that are not characterised by simple sinusoids or polynomials, and hence might have survived the cosine detrending. Of course though, with this we must remain vigilant to avoid removing transits.

We create 48 copies of the cosine-detrended light curve and to each one we independently apply a Tukey's biweight filter. The window length used decreases linearly from the $3 \tau_{\rm min}$ for the 1st light curve to $0.75 \times 3 \tau_{\rm max}$ for the 48th, where $\tau_{\rm min,max}$ are the bounds of the expected transit duration distribution calculated in Sect.~\ref{subsec:detrending_transit_duration_detrending}. The factor of 3 comes from \citet{hippke2019b}; the biweight filter should almost completely preserve a signal with a timespan 1/3rd of the window length. The factor of 0.75 was chosen by us to make the detrending slightly more aggressive, even if some of the longest duration transits might be partially affected. All other parameters in the detrending are set to the \textsc{Wotan} defaults. We then create a single light curve by calculating the expected transit duration at each time step (Eq~\ref{eq:tau}) and then choosing the corresponding detrended light curve from the 48 above. 

\begin{figure*}  
\begin{center}  
\includegraphics[width=0.99\textwidth]{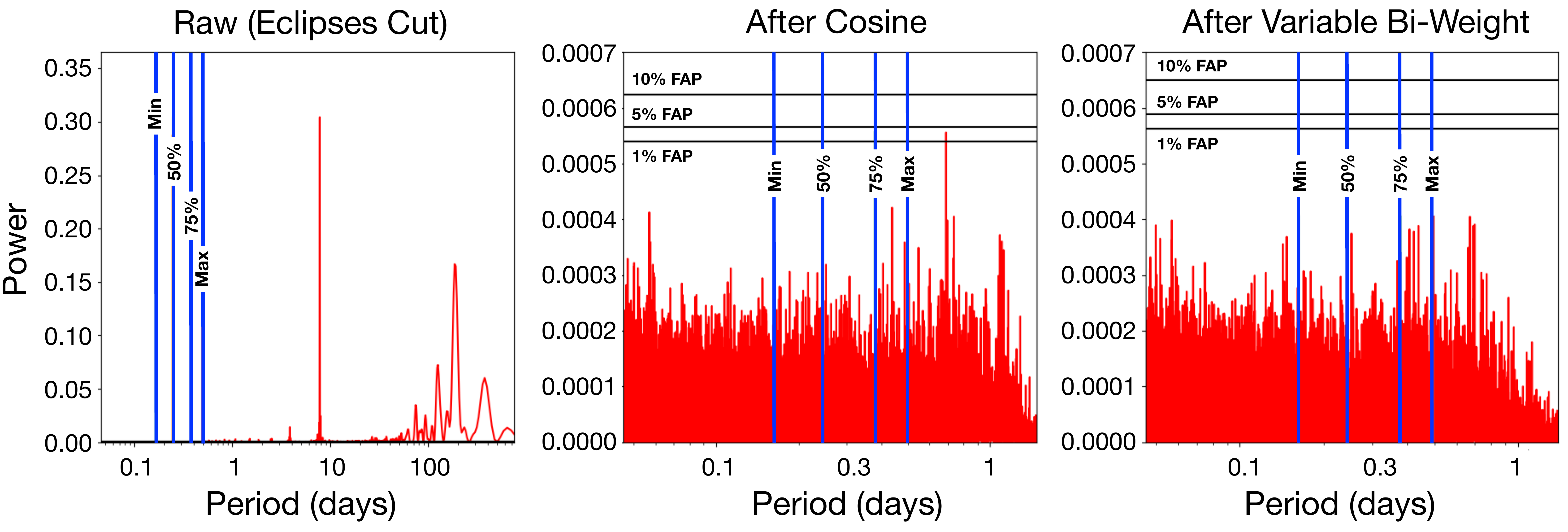}  
\caption{Lomb-Scargle periodogram for the Kepler-47 lightcurve.  The blue vertical lines correspond to the transit durations of a putative planet at $6.1P_{\rm bin}$ ($=45.4$ days), where the transit duration is primarily a function of the binary's orbital phase at transit (see Fig.~\ref{fig:transit_duration}). The minimum and maximum lines corrspond to the shortest and longest possible transits. The 50\% line is a transit duration that is longer than half of all possible transit durations randomised over the binary phase. The 75\% line is calculated similarly. On the left is for the light curve after only the eclipses have been removed, i.e. before any detrending. The largest peak is at 7.78 days, corresponding to the rotation period of the primary star \citep{orosz2012b} and slightly longer than the binary orbital period of 7.45 days. The second peak is at just less than 200 days, most likely corresponding to a trend between quarters of Kepler data. The middle panel is the light curve after applying the cosine detrending filter (Sect.~\ref{subsec:detrending_cosine}). The vertical scale has shrunk significantly because those large spikes of long-period variability have been removed. The horizontal scale is zoomed into shorter periods, where there is now the highest periodogram power. Very little stellar variability remains, except for one spike at 0.7 days which is above a 1\% false alarm probability. The right panel is the light curve after a subsequent application of the variable Tukey's biweight filter, which removes this spike. What remains is a flat (white noise) light curve plus planet transits. }
\label{fig:periodogram}
\end{center}  
\end{figure*}

In Fig.~\ref{fig:periodogram} we  illustrate this procedure for the known circumbinary system Kepler-47. We plot a Lomb-Scargle periodogram of the light curve after applying different layers of the detrending. Initially there  a lot of periodicity in the light curve, most prominently near the binary period at $7.4$ days,  and also at longer periods of $\sim200$ days, most likely corresponding to quarter by quarter variations in the Kepler spacecraft instruments. The cosine filter removes all of this long-timescale power. What remains is a small peak of stellar variability at a timescale of roughly half a day.  This is then removed by the variable biweight filter, leaving us with a flat light curve plus the exoplanet transits.

\subsection{Final cuts from the data}\label{subsec:detrending_finalcuts}

We apply a few final cuts to remove any artifacts from the detrended data, where we define an artifact as any significant deviation from a flat light curve that could not possibly be a transit. This process has a mixture of automation and human intervention. There are four things we look for.

\subsubsection{Gaps in time}\label{subsubsec:detrending_finalcuts_gaps}

A challenge for detrending eclipsing binary data is dealing with frequent gaps in the data from where those eclipses were, in addition to the standard data gaps. The \textsc{Wotan} algorithm typically is very good at treating time gaps; for small gaps the detrender fits the data over the gap without distortion, and for large gaps the detrender creates independent fits to either side of the gap. However, in some exceptional circumstances the detrender creates a poor single fit across a gap in the data. The typical result in the detrended light curve is an upwards or downwards hook, on either one or both sides of the gap in time. 

We detect such artefacts automatically by finding any temporal gap longer than 2 hours and calculating the mean flux for the 2 hours before and 2 hours after the gap. A difference of greater than 4 standard deviations (calculated from the light curve detrended up until this point) is considered significant \citep{hippke2019a}. In this case we cut out all data on the preceding and proceeding 0.8 days of the gap.

\subsubsection{Jumps in flux}\label{subsubsec:detrending_finalcuts_jumps}

In Kepler data a large gap in time may often be associated with a large jump in flux. An example event would be the re-orientation of the spacecraft between quarters,  causing the target to fall onto a different CCD. Some of these undesirable parts of the lightcurves are removed by the \textsc{Lightkurve} quality flags, but not all. Such an event is typically well-handled by \textsc{Wotan}, or as a backup is identified based on the gap in time in Sect.~\ref{subsubsec:detrending_finalcuts_gaps}.

An event {\it not} well-handed by \textsc{Wotan} is an offset in the flux over a short timespan of what is otherwise a smoothly varying light curve. Because there is little gap in time, \textsc{Wotan} attempts detrending with a single fit across the jump. The fit before the jump is distorted downwards, whereas the fit to the right of the jump is distorted upwards. Subtracting this away yields a light curve with differential ``hooks'' on both sides of the data (unlike in Sect.~\ref{subsubsec:detrending_finalcuts_gaps} where there may be a hook on just one side of the data). 

We identify and remove these events automatically. We loop through the light curve and for each timestep we calculate the mean flux over the preceeding and proceeding 2.1 hours. A ``jump'' is identified at a given timestep if the following three conditions are met. First, the mean flux both before and after the time step must be individually more than two standard deviations from the mean flux of the detrended light curve. Second, the absolute difference between the mean flux before and after must be greater than four standard deviations (like in Sect.~\ref{subsubsec:detrending_finalcuts_gaps}). Third, the mean flux before the time step must have a different sign to the mean after the time step, i.e. there must be one upwards hook and one downwards hook. This last step helps us avoid removing short transits.

As in Sect.~\ref{subsubsec:detrending_finalcuts_gaps} we remove the preceding and proceeding 0.8 days of data around any time step identified as a jump.

\subsubsection{Times of low flux common to many different binaries}\label{subsubsec:detrending_finalcuts_commondips}

By this stage in the detrending most non-transit signals have been removed. However, it was discovered that sometimes even transit-looking signals were undesirable. Cross-talk between different Kepler pixels is an instrumental defect which can make transit-like signature appear in multiple different  targets (not just binaries) at the same time. 

\begin{figure*}  
\begin{center}  
\includegraphics[width=0.99\textwidth]{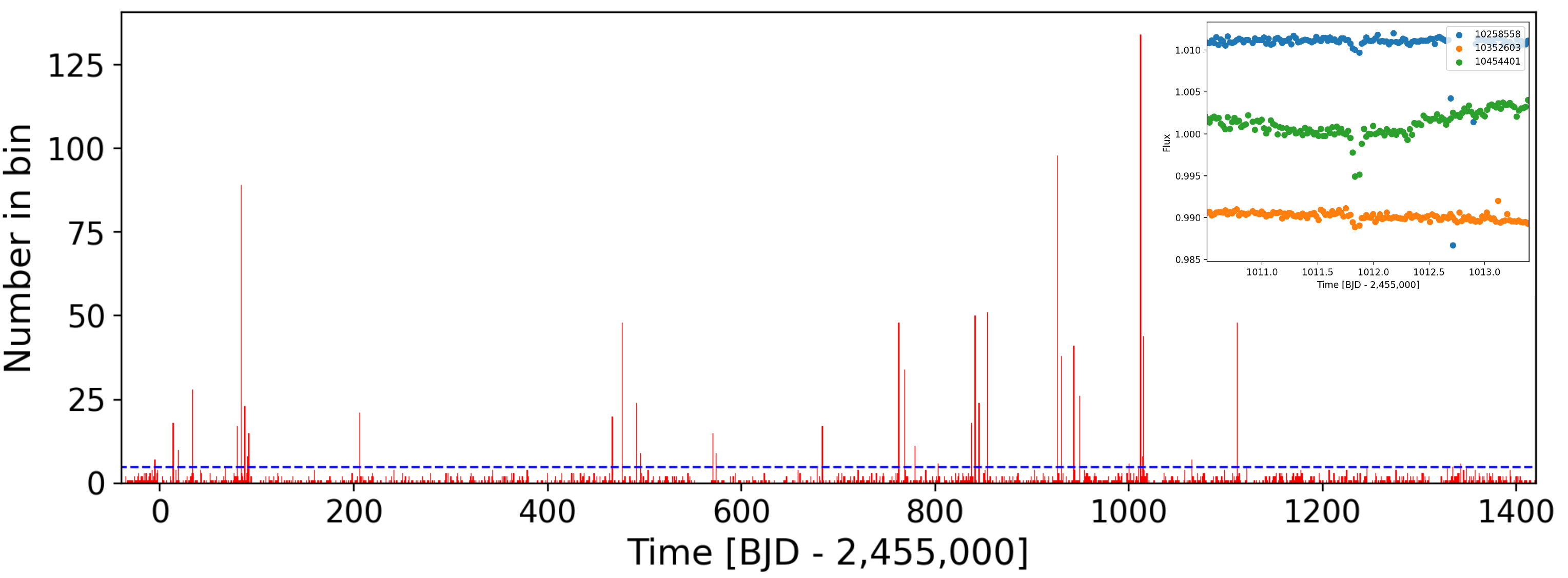}  
\caption{Histogram of light curve times corresponding to the one of the lowest nine flux points in that given binary, after running our code to detrend and remove eclipses. The histogram bin width is one day. The sample comprises 270 eclipsing binaries with periods from 7 to 50 days from the Windemuth catalog \citep{windemuth2019}. There are 35 timestamps that correspond to a deep flux point in more than five of the 270 light curves. Flux at these timestamps above this threshold (blue dashed line) are removed. The inset shows the most commonly-seen feature, at 1011.8 days, in three example light curves. The absolute flux values in this example are arbitrarily offset for clarity. It is transit-like in appearance and is visible in almost half of the tested light curves. Our cautionary list of times corresponding to common false positives is published in the supplementary online materials.}
\label{fig:common_false_positives}
\end{center}  
\end{figure*}

To find such events we ran the detrending algorithm on  270 of the binaries from the Windemuth catalog, with cuts made to only include binaries between 7 and 50 day periods (similar to the known Kepler circumbinary planets,  \citealt{welsh2018,martin2018}). For each binary we outputted the times corresponding to the 9 points of lowest flux, where we avoided listing multiple times within a 1 day bin. We then created a histogram of the deepest points across all these binaries, using day-wide bins, in Fig.~\ref{fig:common_false_positives}.

\begin{figure}  
\begin{center}  
\includegraphics[width=0.49\textwidth]{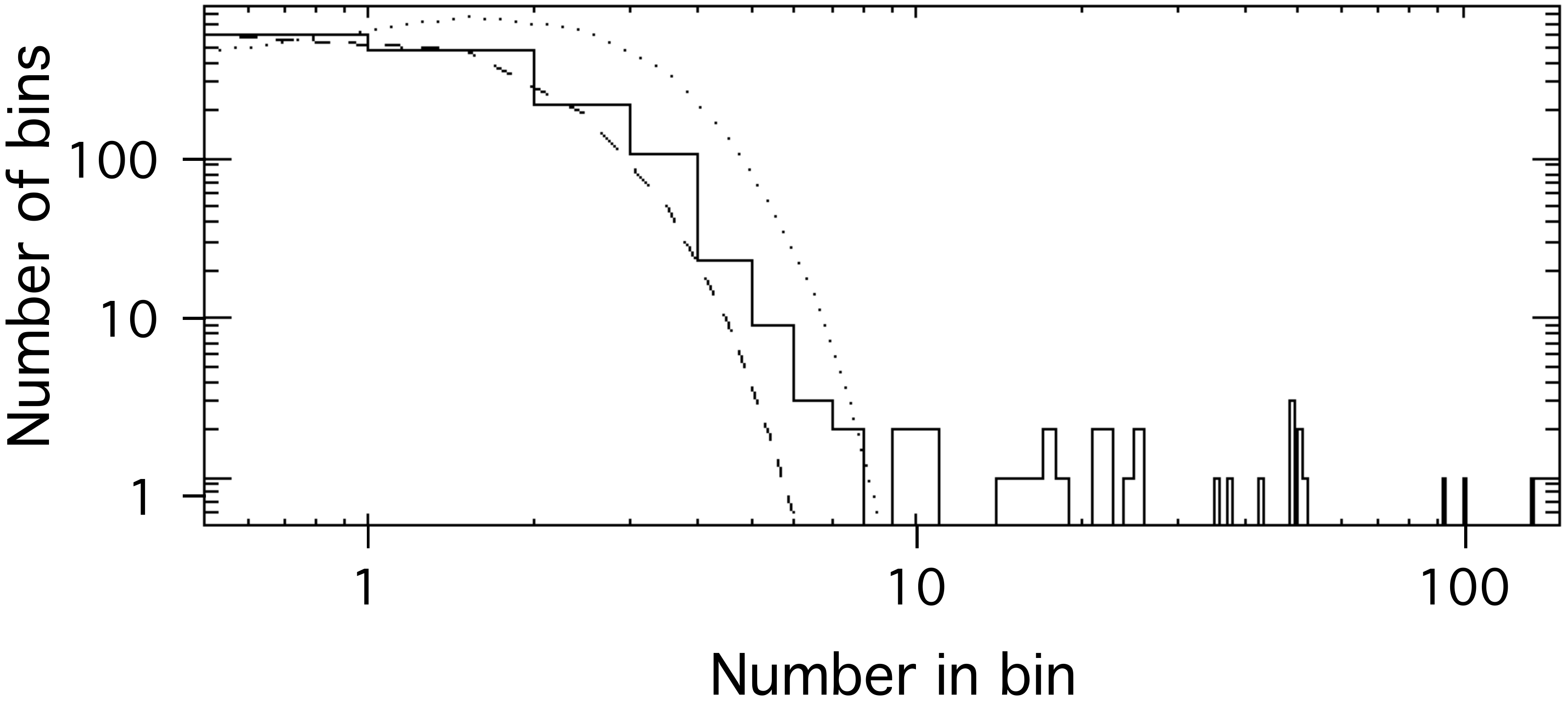}  
\caption{The distribution of Fig.~\ref{fig:common_false_positives}, i.e. the number of systems with low data at each time stamp (actually binned at 1 day).  Poisson distributions are shown for comparison, which assume the events are random in time, as they would be if they are astrophysical. The bin from 0-1 shows the number of 1-day intervals with no low points, the bin from 1-2 corresponds to 1 low point, etc. Over the 1467 bins of daily data, 2430 low points were identified (9 in each of 270 light curves), for a mean rate of $\lambda=1.656$ per bin. The Poisson distribution using this rate, normalized to 2430 low points, is dotted. It is ill-fitting, showing we can rule out that all the hypothesis that all the points are from a random-in-time distribution. Instead, if we suppose that some points are random and others indicate instrumental problems at particular times, then we can use the first two bins to estimate a mean rate of $\lambda=478/599=0.798$ over 1306 low points, which is shown as the dashed Poisson distribution. On this basis, we identify time bins with 5 or more low points  (e.g. 1011.8 days in Fig.~\ref{fig:common_false_positives}) as more likely clumped than random-in-time, hence more likely instrumental than astrophysical.}
\label{fig:poisson}
\end{center}  
\end{figure}

We discovered 35 low flux times that were common in more than 5 of the 270 eclipsing binaries, which we determined to be more likely instrumental than random clumping of real astrophysical events via the analysis of Fig.~\ref{fig:poisson}. In Fig.~\ref{fig:common_false_positives} we demonstrate one event which was seen at time 1011.8 days (BJD 2,455,000) and is transit-like in appearance. It was impressively seen in 134/270 binaries.

We created a list of the 35 low flux times which is published online to warn other planet-hunters. This list is incorporated into the detrending algorithm such that for every target all data is removed at these times. It is possible that by removing $\sim 35$ days of data we remove some legitimate, unique transits of circumbinary planets, however it would be difficult to be convinced of their authenticity if many other binaries had a dip at the same time.

\subsubsection{A final semi-automatic cut}\label{subsubsec:detrending_finalcuts_final}

A final check is made of the light curve for any other events which are clearly spurious. The code automatically produces a list of potential events and then human input is used to decide which (if any) are to be removed. Potential artifacts are produced by the code in two ways. First, the 9 deepest flux points are listed as in Sect.~\ref{subsubsec:detrending_finalcuts_commondips}. Second, any times are listed where there is an eight standard deviation change in flux from one data point to another. Such an instance could correspond to a deep transit (or a less deep transit with a sharp in/egress), which is why the code simply displays these sections of the light curve rather than removing them automatically. A record is made of any sections of the light curve which are manually cut.  When testing \textsc{Stanley} and applying it to the known circumbinary systems we typically perform less than five manual cuts (and often zero). The duration of these manual cuts is typically similar to that of a transit, i.e. hours, and hence representing less than 0.1\% of the total light curve.


\section{Search algorithm}\label{sec:search}

The main aspect of the \textsc{Stanley} code is to identify small transiting planets.   We conduct a brute-force search over a static grid. For every set of parameters (masses, radii and orbital elements) we use an N-body integrator to calculate a mask of transit times and durations over the four year Kepler mission. We then phase-fold the photometric data on these variable transit windows. Finally, we determine if the phase-folded light curve corresponds to a significant dip in the data. 

Unsurprisingly, this method is slow and computationally expensive. We would encourage the application of more sophisticated algorithms such as genetic algorithms and simulated annealing. However, by indiscriminately sampling the entire grid we at least know that the best solution we find will be a global one, not local.

\subsection{Defining the search grid}\label{subsec:search_grid}

Both the binary and planetary orbits have six Keplerian orbital elements: period $T$; eccentricity $e$; argument of periapse $\omega$, inclination $I$, longitude of ascending node $\Omega$ and starting orbital phase, which we characterise with the starting  true longitude $\theta_0$.  We therefore have 12 orbital elements. Both of the stars and the planet have a mass and radius, bringing us to 18 different parameters to define our search grid.

\textsc{Stanley} is written such that a brute-force search over this 18-dimensional grid is  possible. However, for a blind planet search, this is highly unrecommended. In particular, using an N-body algorithm at each point would be prohibitively computationally expensive. We therefore need to have a set value for many of these 18 parameters to reduce the search grid dimensionality, and to optimise the search grid for the remaining parameters.

\subsubsection{Parameters we ignore}\label{subsubsec:search_grid_ignore}

Some of these 18 parameters can be ``ignored'', by which we make an informed simplification of that parameter to a set value. 

\begin{itemize}
    \item $R_{\rm p}=0$.  Our algorithm does not fit transit shapes, but rather stacks together flux within a calculated transit interval. The depth of that flux will of course depend on the planetary radius, but it is not something we fit for or need to know a priori. The width of the interval, which is important for us, is dependent on $R_{\rm p}$, but this is a small factor since $R_{\rm p} \ll R_{\rm A,B}$.
    \item $m_{\rm P}=0$. The mass of the planet affects the dynamics of the 3-body system, which in turn affects the timing of transits and eclipses. However, only in about half of the known circumbinary systems were eclipse timing variations measurable. These eclipse timing variations were caused by gas giants and only on the order of minutes (less than one observing cadence), and hence can be safetly ignored, particularly when searching for low-mass planets. Perturbations on the binary's orbit could also have a back-reaction on the planet's orbit, varying its transit timing, however this may be ignored for detection.
    \item $\Omega_{\rm bin}=0^{\circ}$. When considering transits we are not sensitive to individual values of $\Omega$, but rather just the difference $\Delta\Omega$. It is typical to set $\Omega_{\rm bin}=0^{\circ}$, reducing the dimensionality by one. 
    \item $I_{\rm bin} = I_{\rm p} = 90^{\circ}$ and $\Delta \Omega=0^{\circ}$.  We assume completely flat, edge-on systems. The binary is known to be near $90^{\circ}$ since it eclipses, and any planets will have an inclination near $90^{\circ}$ when transiting. However, this flat assumption can be problematic. A misalignment as small as one degree can affect the transit signature. The 3D orbital orientation of any misaligned planet precesses (a rotation of $\Delta \Omega$). This changes the transit impact parameter $b$, even bringing it above 1, i.e. the planet may stop transiting \citep{schneider1994,martin2014,kostov2014}. By creating a transit mask with a flat system we therefore assume transits on every passing. We therefore risk folding noise from a transit epoch with $b>1$ on top of true transits, diluting the signal. To model and hence try to avoid this effect we would have to expand the search grid dimensionality by two ($I_{\rm p}$ and $\Delta \Omega$), and use a very fine grid to precisely calculate the transit and non-transit epochs. This would be computationally prohibitive. Furthermore, for small circumbinary planets even if you could model the inclination effects it may still be impossible to detect it with significance if lots of its shallow transits are missed. With this in mind, we pursue this flat model and accept we may be less sensitive to misaligned planets. In Sect.~\ref{subsec:results_known} we show that with a flat model we can still re-discover misaligned planets which only have partial transitability (Kepler-47d, -413b and -1661b).
\end{itemize}

\subsubsection{Parameters with  a measured value}\label{subsubsec:search_grid_calculatedvalue}

As described in Sect.~\ref{subsec:detrending_parameters}, the Villanova and Windemuth catalogs combined fully characterize the two binary stars and their Keplerian orbit, i.e. $m_{\rm A}$, $m_{\rm B}$, $R_{\rm A}$, $R_{\rm B}$, $T_{\rm bin}$, $e_{\rm bin}$, $\omega_{\rm bin}$ and $\theta_{\rm 0,bin}$. We only take the best-fitting value of each parameter. It is understood that the photometrically-derived stellar masses from \citet{windemuth2019} are only accurate to $\sim20\%$, and these masses affect the dynamics and transit timing of the planet. However, we will demonstrate the ability to recover the known circumbinary planets despite using \citet{windemuth2019} stellar masses which are known to differ slightly from the true values. \textsc{Stanley} is also set up to search over a range of stellar and binary parameters if desired.


\subsubsection{Parameters we search over}\label{subsubsec:search_grid_search}

The remaining parameters in our blind planet search are $P_{\rm p}$, $e_{\rm p}$, $\omega_{\rm p}$ and $\theta_{\rm 0,p}$. We take care to optimize this grid such that it is both a) as computationally-efficient as possible and b) as unbiased as possible.

We use a grid of planet periods that is neither constant in linear or log space. We instead optimize it based on the expected transit duration, similar to what \citet{ofir2014} did for single stars. The planet period grid spacing is described by $\delta P_{\rm p}(P_{\rm p})$, i.e. the step size varies with $P_{\rm p}$. To calculate $\delta T(P_{\rm p})$, we first calculate the relative velocity between the primary star and planet using Eq.~\ref{eq:velocity}. The shortest possible transit duration is  

\begin{equation}
   \tau_{\rm min}(P_{\rm p}) = \frac{R_{\rm A}}{|V_{\rm p}(P_{\rm p})-V_{\rm A}|},
\end{equation}
corresponding to the planet and primary star moving in opposite directions with respect to the observer at transit. We set $\delta P_{\rm p} = 3\tau_{\rm min}$. Our period grid is therefore derived with the precision to hit even the smallest duration transit  within a factor of three.  The factor of three is because the code ultimately allows slight deviations of the transit times from the N-body calculations up to three times the predicted duration. This small ``sliding'' effect is largely to account for imprecise measured stellar parameters and is further described in Sect.~\ref{subsec:search_transitmask}.

The minimum period searched  corresponds to a semi-major axis of

\begin{equation}
    \label{eq:stability_limit}
    a_{\rm p,min} = 2.2 a_{\rm bin}(1+e_{\rm bin}),
\end{equation}
 where interior orbits would be almost guaranteed to be unstable \citep{dvorak1984,holman1999,mardling2001,quarles2018}. The longest period searched is two years, since we demand at least three primary transits within the four years of Kepler data. This will inhibit detection of planets such as Kepler-1647b, which passed the binary twice during the four years of Kepler \citep{kostov2016}. This is a gas giant though, and we imagine that finding Earth-like circumbinary planets with only  two transits would be nearly impossible.  In Sect.~\ref{subsubsec:results_known_kepler1647} we loosen the detection requirements and demonstrate we can then recover Kepler-1647b.


The spacing in the starting planet phase, $\theta_{\rm 0,p}$ also varies with the planet period $P_{\rm p}$, and is calculated by 
\begin{equation}
\delta \theta_{\rm 0,p}(P_{\rm p}) = 3\times360^{\circ}\times\frac{\tau_{\rm min}(P_{\rm p})}{P_{\rm p}}.
\end{equation}
The $\theta_{\rm 0,p}$ grid becomes finer at greater periods. 

For the planet eccentricity and apsidal alignment traditional search methods either have uniform grids in $e$ and $\omega$ or uniform grids in $e\sin\omega$ and $e\cos\omega$ (or possibly a slight variation such as $\sqrt{e}$ instead of $e$). We propose that both are problematic. 

If $\delta e$ and $\delta \omega$ are constant then the effects of a large eccentricity are relatively under-sampled (and those of a small eccentricity or even circular orbit are over-sampled). On the other hand, whilst using a square grid of $\delta(e\sin\omega)$ and  $\delta(e\cos\omega)$ does avoid that problem, it means that the range of $e$ sampled varies with $\omega$ varies by a factor of $\sqrt{2}$\footnote{ For a square grid of $\delta(e\sin\omega)$ and  $\delta(e\cos\omega)$ the distance from the origin to the grid point is equal to the eccentricity. At say $\omega=45^{\circ}$, i.e. in the top right corner of the square grid, the maximum eccentricity value will be $\sqrt{2}$ times that at $\omega=0^{\circ}$ along the horizontal axis of the grid.}

We apply a novel technique of a circular grid in $e\cos\omega$ and $e\sin\omega$. We have constant steps in eccentricity out to a maximum value that is independent of $\omega$. For a given eccentricity we then define $\delta\omega=\delta e/e$ in radians, which is the equivalent to stepping around the circumfrence of a circle in $e\cos\omega$ vs $e\sin\omega$ space, where the step size along the circumfrence is the same for every $e$. 

For our search we use $\delta e=0.067$ between $e=0$ and $e=0.2$. This leads to  four steps of eccentricity and 34 different combinations of $(e,\omega)$, where higher eccentricities having more steps in $\omega$ according to our circular grid. Our range in eccentricity covers all of the known planets, where all but one has a small eccentricity less than 0.1. The grid spacing $\delta e=0.067$ is a somewhat arbitrary choice, but was found to  re-discover the known circumbinary planets. A finer grid would significantly increase computation time. On the other hand, the assumption of circular orbits was found to be ineffective, even for known planets with very small eccentricities. We believe the cause to be a non-negligible amount of apsidal advance over four years of {\it Kepler} observations causing the transit times to shift, even for the most circular orbits. Three-body interactions also prevent planets from maintaining perfectly circular orbits.

Overall, we typically use between 10 and 100 million individual transit models when searching for a Kepler circumbinary planet around any given target.

\subsection{Running the N-body simulation}\label{subsec:search_nbody}

For a given lightcurve we run the {\sc Rebound} IAS-15 N-body integrator \citep{rein2012,rein2015} on each set of parameters in the search grid to produce a set of transit times and durations. This is the slowest part of the algorithm, and is often best suited to a computing cluster, which is covered in Sect.~\ref{subsec:search_cluster}. 

Each N-body integration is run over the entire timespan of the light curve, which is typically about four years for {\it Kepler}. Transit times are calculated using an iterative procedure, where the code notes a sign change in the difference in the horizontal positions between the planet and primary star, and then steps backwards and forwards with increasingly small steps until the solution is converged upon. This is the default means of calculating TTVs in {\sc Rebound}\footnote{\url{https://rebound.readthedocs.io/en/latest/ipython/TransitTimingVariations.html}}. Note that we ignore light travel time effects, which should be no more than seconds in our examples \citep{welsh2012} and hence can be ignored for our purposes. We only calculate transits on the primary, and discuss this limitation in Sect.~\ref{subsec:discussion_improvements}.

The transit duration is then calculated as in Eq.~\ref{eq:tau} based on the relative velocities of the primary star and planet, and the size of the primary star. Recall that we assume edge-on, coplanar systems so $b=0$, and the planet radius is considered negligible and hence set to zero.

Even though a stability limit criterion was invoked in creating the range of planet periods tested  (Eq.~\ref{eq:stability_limit}), such an equation is simply an approximation and does not take into account finer complexities such as the relative planet and binary phases, or structures in the stability map such as mean motion resonances. Consequently, some tested planets may still be unstable. We invoke a very simple test for stability in the N-body integration. At 50 equally spaced timesteps along the integration we calculate the planet's osculating Keplerian eccentricity and compare it to its starting value. If it varies by more than 0.1 we consider this system to be unstable and stop the integration. 

We emphasise that this is a very rough instability approximation. In a four year integration we are typically not seeing instability fully unfold (i.e. the planet being ejected) but rather a change of 0.1 is seen as a sign that instability may come. A true test of instability would involve integrations for thousands or even millions of years (e.g. \citealt{quarles2018}), which is not feasible with an N-body brute-force search algorithm. We recall that the purpose of this algorithm is an initial identification of transiting circumbinary planets, and any candidates found would be expected to receive further scrutiny, including of their long-term stability.


\subsection{Fitting the transit mask}\label{subsec:search_transitmask}

\begin{figure*}  
\begin{center}  
\includegraphics[width=0.99\textwidth]{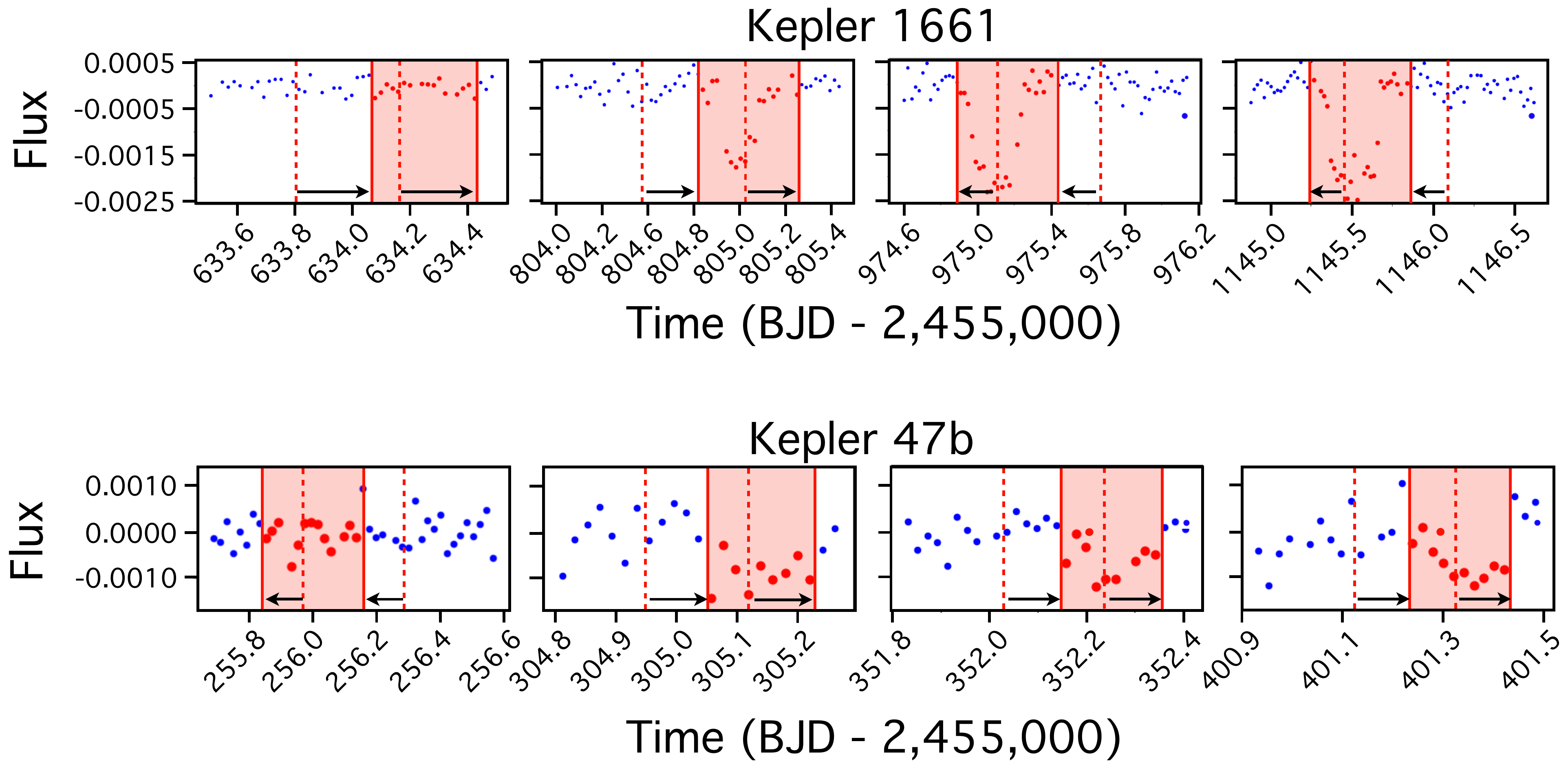}  
\caption{ Example transit detection for four transits of the $3.87R_{\oplus}$ Kepler-1661b (top) and the $3.0R_{\oplus}$ Kepler-47b (bottom). The initial N-body fit of the transit time and duration is demarcated by the red dashed lines. Each transit window is equal to three times this duration, and hence  some windows are longer than others. The red shaded region bounded by solid lines is the final transit picked out by the algorithm, which is selected within this window to minimize the in-transit flux.  For Kepler-1661 the planet did not transit for the first half of the Kepler mission due to a slight but non-negligible misalignment of $\sim1^{\circ}$. In the first panel this is the epoch where {\sc Stanley} predicts a coplanar planet would have transited. For the later three epochs the planet has precessed into transitability.}
\label{fig:transit_slider_plot}
\end{center}  
\end{figure*} 

For every set of circumbinary parameters the N-body code outputs a sequence of transit times and durations, which we then fit to the light curve. If our search grid contained an exact corresponding set of parameters then the N-body model would produce a perfect fit. In practice though, this is unlikely for two reasons. First, our stellar parameters are only precise to $\sim20\%$ \citep{windemuth2019}. Second, computing time prohibits constructing a grid of planet parameters sufficiently fine to produce transit times precise to sub-observing cadence ($<30$ minutes). 

To account for any slight imperfections in the N-body model, the algorithm loops through each of the transits and creates a window of the light curve centred on the N-body transit time, where the width of this window is three times the N-body transit duration ($3\tau$, Eq.~\ref{eq:tau}). We then slide a smaller window of width $\tau$ to find the time corresponding to the lowest mean in-transit flux. We then note this as the transit time. We do not vary the transit duration.

For each transit we take the in-transit flux data and subtract from it the mean out-of-transit flux (within the $3\tau$ window). We then add these relative transit fluxes to a ``global transit'', i.e. a circumbinary transit which has been phase-folded on a variable transit interval with variable transit durations. By looking at a relative flux between inside the transit and slightly either side of it, as opposed to an absolute in-transit flux, we can account for any remaining stellar or instrumental variations which were not removed by the detrending process.




The SNR of this global transit is calculated to be

\begin{equation}
\label{eq:SNR_global_transit}
    {\rm SNR}=\frac{\bar{f}\sqrt{N}}{s_{\rm oot}},
\end{equation}
where $\bar{f}$ is the mean in-transit relative flux for the global transit, $N$ is the number of in-transit data points and $s_{\rm oot}$ is the standard deviation calculated over the entire light curve excluding the in-transit data.

We illustrate this procedure in Fig.~\ref{fig:transit_slider_plot} for four transits of Kepler-1661 and Kepler-47. The N-body  code initially predicts a transit time and duration centered in each window, and then we slide a transit mask along the window to find the minimum flux. In these eight transits the initial transit time is about half a transit duration off the true centre. These corrections are on the order of  $\sim 2-4$ hours, which is small compared to the TTVs on the order of $\sim16$ hours and $\sim50$ hours for Kepler-1661 and Kepler-47b, respectively.

These two sets of transits are chosen to demonstrate  some idiosyncrasies which the algorithm sometimes has to deal with. For Kepler-1661, the planet is mutually inclined by $\sim1^{\circ}$ with respect to the binary. Consequently, the planet initially does not transit until the second half of the Kepler mission, and then the transits increase in depth as the impact parameter decreases. As discussed in Sect.~\ref{subsubsec:search_grid_ignore}, we assume coplanar orbits since scanning over a grid of 3D orientations would be computationally prohibitive. Here we see the drawbacks of this: we fold a non-existent transit near  634 days (BJD - 2,455,000), and also we over-estimate the transit durations.

For Kepler-47b Fig.~\ref{fig:transit_slider_plot} shows evident (but shallow) dips for this $3R_{\oplus}$ planet in the second, third and fourth panels. In the first panel, the light curve looks flat. In fact, a transit does exist near 256 days, but it was unfortunately detrended away as seen earlier in Fig.~\ref{fig:detrending_example}. This planet transits 24 times and so the transit timing model is sufficiently constraining to predict a transit here.

Despite these small challenges, Sect.~\ref{subsec:results_known} shows that both planets are detected with high significance.


We note here a caution for the sliding method. Take one extreme, where we use solely the N-body algorithm to predict transit times and do no sliding. Since these times are entirely based on a circumbinary planet model, then if they line up with actual dips in the light curve we would be rather inclined to believe a planet has been found. However, as illustrated in Fig.~\ref{fig:transit_slider_plot} the N-body transit times can be slightly off even for known, legitimate planets, and hence such an algorithm would be inefficient. At the other extreme, one could completely avoid using an N-body model to predict transit times and simply scan through the light curve looking for the deepest flux points\footnote{after removing eclipses and applying the detrending}. This would be very efficient, and indeed guaranteed to find all of the dips in the lightcurve. However, there would be no physical model for the timing of these dips, and hence we would be less convinced we had found a planet and not just artifacts of poor detrending. Indeed the QATS algorithm follows this method to an extent,  and is thought to suffer from this for circumbinary planets (\citealt{orosz2012b} and see Sect.~\ref{subsubsec:discussion_comparison_qats}).


\subsection{Applying quality cuts}\label{subsec:search_quality_cuts}

Several cuts are made by \textsc{Stanley} to reduce the likelihood of false positive solutions. First, when fitting each N-body transit model we exclude any transits for which more than 25\% of the data is absent (based on the expected number of data points for a given transit duration). We then also do this on a global scale for the entire transit model, and demand that at least 50\% of the expected data points are in the light curve. 

These cuts avoid the algorithm artificially finding low mean flux solutions that are driven by one or two very deep outlier events and then the majority of the transit mask falls in gaps, and hence not affecting the mean flux. Such situations would likely be false positives, or at best true planet transits that we would never be able to confirm with the existing data. 

 The second criterion is that we require there to be at least three detected transits that are roughly consistent. Specifically,  we calculate the SNR for each individual transit, determine the maximum value SNR$_{\rm max}$, and demand that at two of the other transits have a SNR greater than $0.45\times{\rm SNR_{\rm max}}$. Any transits that fall in a gap are ignored for the sake of consistency. 

By requiring  at least three roughly consistent transits we avoid selecting a solution which is solely driven by a small amount of deep events, combined with much larger number of insignificant transits. Such inconsistent events would be likely due to a completely non-periodic phenomena such as flares or artifacts in the light curve detrending. Alternatively, they could be produced by real transiting circumbinary planets, but with periods so long that they fall out of our search grid and would be hard to confirm anyway, or at least require the  differing methodology of \citet{kostov2020b} to confirm them. Whilst our results typically use this three transit criterion, in Sect.~\ref{subsubsec:results_known_kepler1647} we show that by loosening it we can recover the 1108-day Kepler-1647b.

 We emphasise that the three transits only need to be {\it roughly} consistent. The $0.45$ SNR factor is somewhat arbitrary but allows for variations in transit duration and depth (and hence SNR) that are common in circumbinary planets, and works to recover the known planets. Loosening how consistent the transits need to be may allow for more marginal detections, for example with grazing transits, and this criterion will be easily editable in the code for the user's preference.


\subsection{The best fitting solution and the signal detection efficiency (SDE)}\label{subsec:search_sde}

We construct a variant of the commonly-used Signal Detection Efficiency (SDE)\footnote{Within transit searches, often referred to as the BLS (Box Least Squares) statistic based on \citet{kovacs2002}} to pick out the best solution and inform us of how significant it is compared with other tested transit models.

First, at each tested planet period we marginalise over all tested orbital parameters (which, in our work is typically $\theta_{\rm 0,p}$, $e_{\rm p}$ and $\omega_{\rm p}$) to find the transit model with the maximum signal to noise ratio of each fit SNR$(P_{\rm p})$.


There is typically a decreasing trend in SNR as a function of $P_{\rm p}$. This is because the number of in-transit points $N_{\rm t}$ decreases as longer period planets transit less \footnote{For planets around single stars, for wider orbits the transits are of longer duration, partially offsetting this reduction in SNR. For circumbinary planets this effect will be true on average, but the binary phase at transit has a much greater effect on the transit duration than the planet period.}. Following \citet{hippke2019a} we remove this trend to avoid (to the best of our ability) a bias towards short-period planets. The Tukey's biweight filter in \textsc{Wotan} \citep{hippke2019b} is employed with a window length of

\begin{equation}
\label{eq:SDE_detrending_window}
    \xi = 10P_{\rm bin}/\pi
\end{equation}
This window length is derived based on the analytic TTV calculations of \citet{armstrong2013} and should be $\sim 4$ times the width of the planet-induced peaks in the SDE curve (in Fig.~\ref{fig:SDE_plots_WITHtransits}).

\begin{figure}  
\begin{center}  
\includegraphics[width=0.40\textwidth]{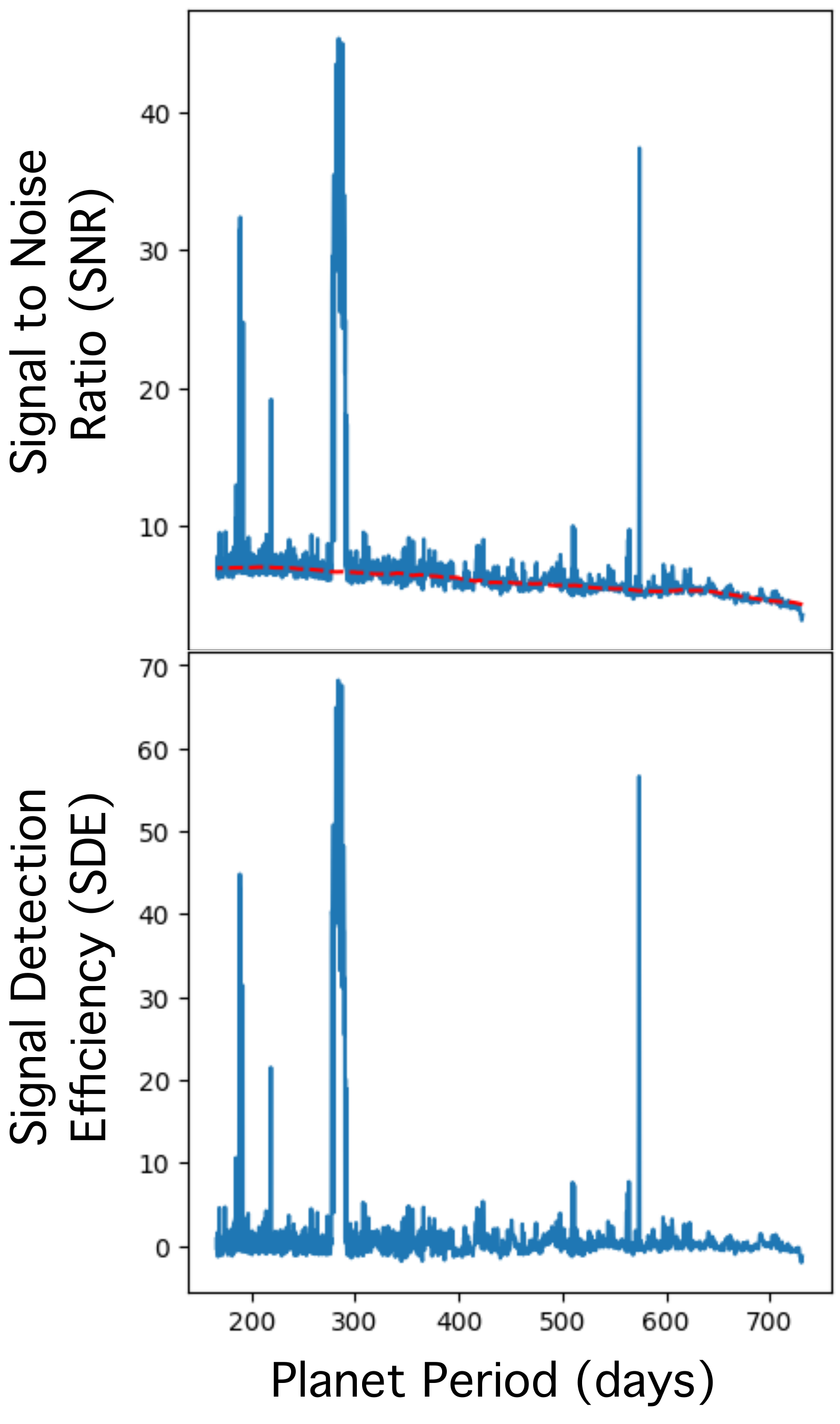}
\caption{Example of the Signal to Noise Ratio (SNR) from Eq.~\ref{eq:SNR_global_transit} for Kepler-34 in the top panel, and the conversion to Signal Detection Efficiency (SDE) from Eq.~\ref{eq:SDE} in the bottom panel. The red dashed line is the trend removed using a Tukey's biweight filter in \textsc{Wotan}, since shorter planet periods will systematically have higher SNRs since more data points are captured in transit.}  
\label{fig:SNR_to_SDE}
\end{center}  
\end{figure}

The SDE$(P_{\rm p})$ is then defined at each planet period  according to \citet{hippke2019b}  as

\begin{equation}
\label{eq:SDE}
    {\rm SDE}(P_{\rm p}) = \frac{{\rm SNR}(P_{\rm p})}{s_{\rm SNR}},
\end{equation}
where $s_{\rm SNR}$ is the standard deviation of  the function SNR$(P_{\rm p})$. This standard deviation is calculated in rolling segments (with width $\xi$, Eq.~\ref{eq:SDE_detrending_window}) of ${\rm SNR}(P_{\rm p})$ and then we take the median value of all of the standard deviation calculations. This means that in calculating $s_{\rm SNR}$ we avoid including any spikes in SNR due to transiting planets. In Fig.~\ref{fig:SNR_to_SDE} we illustrate both SNR$(P_{\rm p})$ and SDE$(P_{\rm p})$ for Kepler-34.

The best-fitting period is the one with the highest SDE. The values for $\theta_{\rm p}$, $e_{\rm p}$ and $\omega_{\rm p}$ are those with the highest SNR (Eq.~\ref{eq:SNR_global_transit}) at this period. Note that the best fitting solution works in the same way if the search grid is expanded to include any of the other orbital and/or stellar parameters.

\citet{hippke2019a} note that independent studies have derived different thresholds for what is considered a reliable detection based on SDE: 7 \citep{siverd2012}; 6 \citep{dressing2015}; 6-8 \citep{pope2016}; 6.5 \citep{livingston2018}; 10 \citep{wells2018}. Ultimately a lower SDE threshold will increase survey completeness and provide a greater number of candidates to follow-up, but with the risk of added false positives. This will be quantified more thoroughly in our second paper when applying \textsc{Stanley} to a larger sample  in the search for small planets.

\subsection{Computation times}\label{subsec:search_cluster}

\textsc{Stanley} can be run on either a single computer or on a computing cluster. The code is set up to split the search grid over $P_{\rm p}$, which typically has the most steps, and run in parallel. The results are ultimately combined seamlessly. To recover the known planets over a period grid spanning up to two years takes typically a couple of 100 computational hours, and so for a blind search a computing cluster\footnote{or a good book} is highly recommended. Given typically 100+ million independent N-body simulations are run for each planet search, this time is not surprising. Reducing the longest tested period can significantly reduce computing time (advice we follow in Sect.~\ref{subsec:results_detection_limits}) because the longest planet periods require the finest grid in $\theta_{\rm p}$. Fitting circular planets would also expidite the fit, but in our experience a circular orbit often failed to fit most of the transits, no matter how fine the grid in $P_{\rm p}$ and $\theta_{\rm p,0}$ was, although we have not strenuously benchmarked this.

We refer the reader to \citet{rein2012,rein2015} for details on the {\sc Rebound} IAS-15 N-body integrator. We note here that it has an adaptive timestep which is primarily set by the shortest orbital period, i.e. of the binary. A given simulation of the 7.4-day Kepler-47 binary will therefore be slower than the 41.1-day Kepler-16. However, our adaptive grid is more dense for longer period binaries, which somewhat evens out the  overall computing time.

\begin{table*}[]

\begin{flushleft}
\footnotesize

\begin{tabular}{llllllllllllllllll}
\hline
Name        & KIC      & $m_{\rm A}$ & $m_{\rm B}$ & $R_{\rm A}$ & $R_{\rm B}$ & $P_{\rm bin}$ & $e_{\rm bin}$ & $\omega_{\rm bin}$ & $t_0$        \\
            &          & $(m_{\odot})$ & $(m_{\odot})$ & $(R_{\odot})$ & $(R_{\odot})$ & (days)          &               & (deg)                & (BJD)          \\
\hline
Kepler-16   & 12644769 & 0.6002      & 0.1913      & 0.6092      & 0.2161      & 41.0768       & 0.1913        & 261.608            & 54965.657634  \\
Kepler-34   & 08572936 & 1.1039      & 1.0779      & 1.1985      & 1.1367      & 27.7953       & 0.4998        & 67.7074            & 54979.723069  \\
Kepler-35   & 09837578 & 1.2161      & 1.0655      & 1.1361      & 0.9515      & 20.7334       & 0.0731        & 82.9675            & 54965.845830  \\
Kepler-38   & 06762829 & 1.0913      & 0.2883      & 1.8608      & 0.2952      & 18.7949       & 0.0038        & 145.2592           & 54971.667903  \\
Kepler-47   & 10020423 & 0.8936      & 0.3341      & 0.9120      & 0.3322      & 7.4482        & 0.0241        & 215.3993           & 54970.693646  \\
Kepler-64   & 04862625 & 1.328       & 0.3831      & 1.6991      & 0.3730      & 19.9999       & 0.1907        & 152.135            & 54967.819337  \\
Kepler-413  & 12351927 & 0.8298      & 0.6051      & 0.7310      & 0.6652      & 10.116        & 0.0079        & 220.6671           & 54972.981520  \\
Kepler-453  & 09632895 & 0.7776      & 0.1831      & 0.7779      & 0.2043      & 27.3216       & 0.0569        & 263.823            & 54965.424466  \\
Kepler-1647 & 05473556 & 1.206       & 0.971       & 1.7773      & 0.9694      & 11.2586       & 0.1486        & 303.7721           & 54956.478952  \\
Kepler-1661 & 06504534 & 0.818       & 0.3656      & 0.7169      & 0.4228      & 28.1621       & 0.094         & 344.9632           & 54976.717044  \\
\hline
Name        & $P_{\rm p}$ range      & $P_{\rm p}$ & $e_{\rm p}$ & $\omega_{\rm p}$ & $\theta_{\rm p}$ & SDE     & Primary       & Primary  &         \\
            &        (days)  & (days)        &             & (deg)              & (deg)         &         & transits (real) & transits (found)      &                \\

\hline
Kepler-16   & [160,730] & 231.3447    & 0.1333         & 261.8182         & 31.3575     & {\bf 14637.5} & 7                & 7   &                                  \\
Kepler-34   & [167,730] & 283.9332    & 0.2      & 63.5294          & 69.0566    & {\bf 68.2}    & 5                & 5         &                 \\
Kepler-35   & [75,730] & 132.429    & 0.1333      & 98.1818            & 35.7593    & {\bf 66.0}    & 6                & 6      &                  \\
Kepler-38   & [62,730] & 105.3419    & 0.2      & 105.8824         & 54.3956    & {\bf 11.8}    & 13               & 10   &                    \\
Kepler-47b   & [25,730] & 48.8588     & 0.0667         & 0.0              & 11.4286    & {\bf 16.1}    & 24               & 22     &                      \\
Kepler-47d   & [25,730] & 187.1520     & 0.1333         & 98.1818              & 96.7619    & {\bf 15.1}    & 6               & 6   &                        \\
Kepler-47c   & [25,730] & 303.3542     & 0.2         & 84.7059             & 64.7619    & {\bf 15.2}    & 4               & 3       &                 \\
Kepler-64   & [85,730] & 140.0077    & 0.1333      & 196.3636         & 50.9589    & {\bf 19.8}    & 10               & 8          &                   \\
Kepler-413  & [33,730] & 65.5662     & 0.2      & 317.6471         & 36.7347    & {\bf 14.9}    & 9                & 8       &                 \\
Kepler-453  & [97,730] & 239.6598    & 0.2      & 127.0588          & 129.8028       & {\bf 361.5}   & 3                & 3       &                  \\
Kepler-1647 & [45,730] & 134.7954          & 0.2          & 127.0588               & 58.9271          & 7.5     & 1                & 0       &                   \\ 
Kepler-1647* & [45,1461] & 1108.2609          & 0.2          & 63.5294               & 54.0267          & {\bf 32.5}     & 2 (Secondary)                & 2 (Secondary)       &                   \\ 
Kepler-1661 & [105,730] & 175.3048    & 0.1333         & 65.4545          & 43.9286    & {\bf 45.0}    & 3                & 3 & \\

\end{tabular}
\caption{ Circumbinary planet parameters for the re-discovery of the 12 known circumbinary planets. Stellar parameters are taken from \citet{windemuth2019} and binary orbital parameters from the Villanova catalog (\url{http://keplerebs.villanova.edu/}). Planet parameters are those fitted by the \textsc{Stanley} algorithm. All parameters are calculated at the specified $t_0$, which always corresponds to a primary eclipse of the binary, i.e. $\theta_{\rm bin}=90^{\circ}$.   Values are shown using standard detection criteria (three roughly consistent transits on the primary star). For Kepler-1647 we also show the results from Sect.~\ref{subsubsec:results_known_kepler1647} where we change the criteria to only two transits on the secondary star, and we accordingly search over a period range up to four years. These results are denoted with an asterisk. At 1108 days this matches the true period, but with \textsc{Stanley} alone we cannot be sure the period is not half this value. Note that this table should  not be used as a reference for these parameters and the reader should instead refer to the discovery papers.  Ultimately, a significant detection  (SDE$\gtrsim8$) can be made for all of the planets.}
\label{tab:known_planets}
\end{flushleft}
\end{table*}

\section{Results}\label{sec:results}


\subsection{Recovering known circumbinary planets}\label{subsec:results_known}

\subsubsection{Main results}\label{subsubsec:results_known_main}

There are ten published Kepler circumbinary systems. We run the \textsc{Stanley} detrending and detection algorithm on all of them and catalogue our fits in Table~\ref{tab:known_planets}. All Kepler systems yield a significant detection with  an SDE above 12, except for Kepler-1647 as expected, since its 1108-day planet does not transit the mandatory three times  (but see Sect.~\ref{subsubsec:results_known_kepler1647}). In  Table~\ref{tab:known_planets} we show our fitted parameters to each system and the detection significance.

\begin{figure*}  
\begin{center}  
\includegraphics[width=0.99\textwidth]{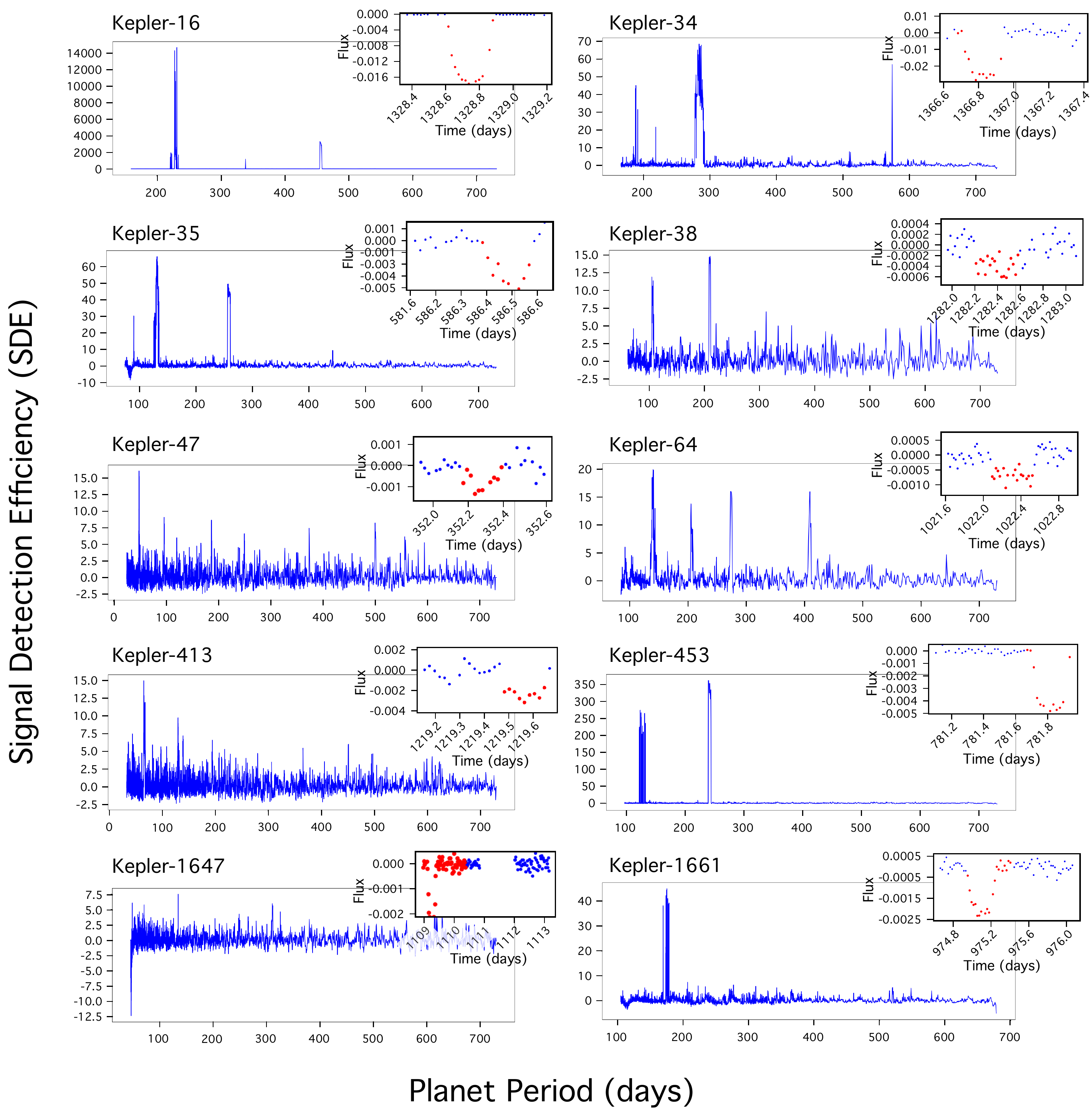}  
\caption{Signal Detection Efficiency (SDE) for the re-discovery of all ten known circumbinary planet systems. Inset in each plot is a representative transit as discovered by \textsc{Stanley}, with red in-transit data points and blue out-of-transit points.}
\label{fig:SDE_plots_WITHtransits}
\end{center}  
\end{figure*}

In Fig.~\ref{fig:SDE_plots_WITHtransits} we show the SDE for each system. A  few of features stand out. For all but Kepler-1647 there is at least one prominent spike, significantly above an SDE of 8. It is noticeable that this spike is rather broad. The main cause of this breadth is that the SDE at a given period is the maximum value marginalised over all different values of $\theta_{\rm p}$, $e_{\rm p}$ and $\omega_{\rm p}$. With these degrees of freedom, even if the planet's period is changed by a few per cent, a significantly different $\theta_{\rm p}$, $e_{\rm p}$ and $\omega_{\rm p}$ could contort the transit timing to still fit the data, and hence provide a similarly good SDE. The non-Keplerian perturbation on the planet's orbit by the binary also ``smear'' the best-fitting solution, as our solutions only indicate the osculating orbital elements at time $t_0$, and they will change significantly over four years.

Another feature seen in the SDE plots are secondary peaks at alias periods of the true peak. Such peaks correspond to solutions where either a sub-set of the transits are found, or all of the transits are found, in addition to some noise.

Kepler-38 is the only case where its true period of 105 days has a significant SDE of 12.2, but it is smaller than the SDE = 14.7 at double the period. This occurs because the planet-binary period ratio is close to 5.5, and the planet's starting phase awkwardly causes its early transits to alternatively be near the primary and secondary eclipses of the binary. Its second, third and fourth transits are all blended with eclipses and hence will not be detected. Transits near secondary eclipse, even if not blended, are of maximum duration ($\theta_{\rm A}=90^{\circ}$ in Fig.~\ref{fig:transit_duration}) and hence are at risk of being flattened by the detrending process. By visually comparing the transit fits at 105 and 210 days it is clear that the 105 day model does include additional, legitimate transits and hence should be favoured. This is an example where some manual intervention may be needed in what is otherwise an automated process. Overall the model predicts 12 out of 13 primary transits for  Kepler-38, even though some of these transits were removing by the detrending algorithm. This demonstrates that circumbinary models are highly constraining if you are able to find them.

For each re-discovered planet in Fig.~\ref{fig:SDE_plots_WITHtransits} we show an example transit as discovered by the algorithm, i.e. with red in-transit points and blue out-of-transit points.

\subsubsection{Multi-planet Kepler-47}\label{subsubsec:results_known_kepler47}

\begin{figure*}  
\begin{center}  
\includegraphics[width=0.90\textwidth]{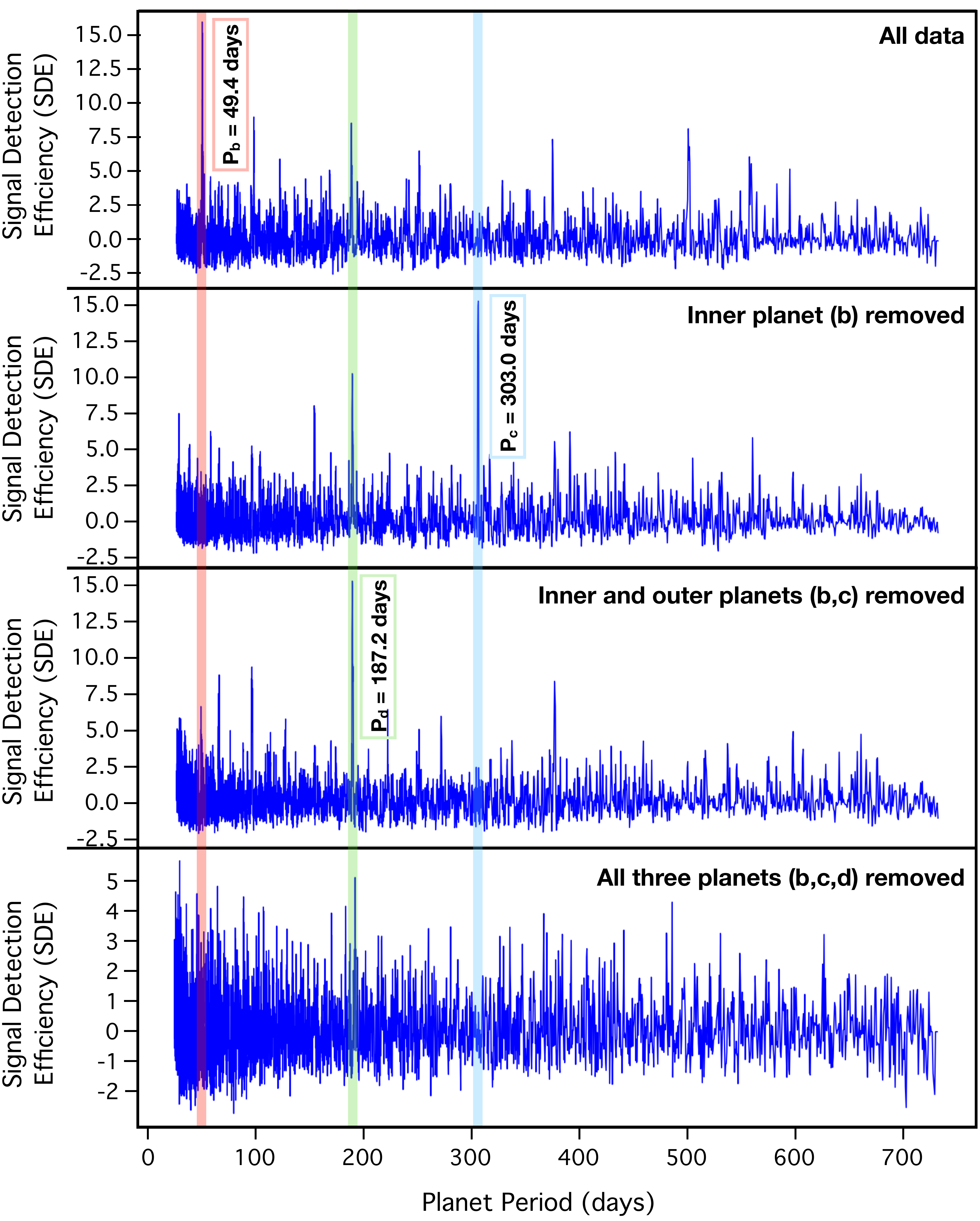}
\caption{SDE for Kepler-47 where we iteratively remove the transits corresponding to the highest SDE. The top panel shows the fit using all of the data. This process shows that we can recover all three planets with a significant SDE$\sim 15$. We also see spikes from aliases of each planet period, which all disappear when the transits have been removed.}  
\label{fig:kepler_47_SDEs}
\end{center}  
\end{figure*} 

\begin{figure*}  
\begin{center}  
\includegraphics[width=0.90\textwidth]{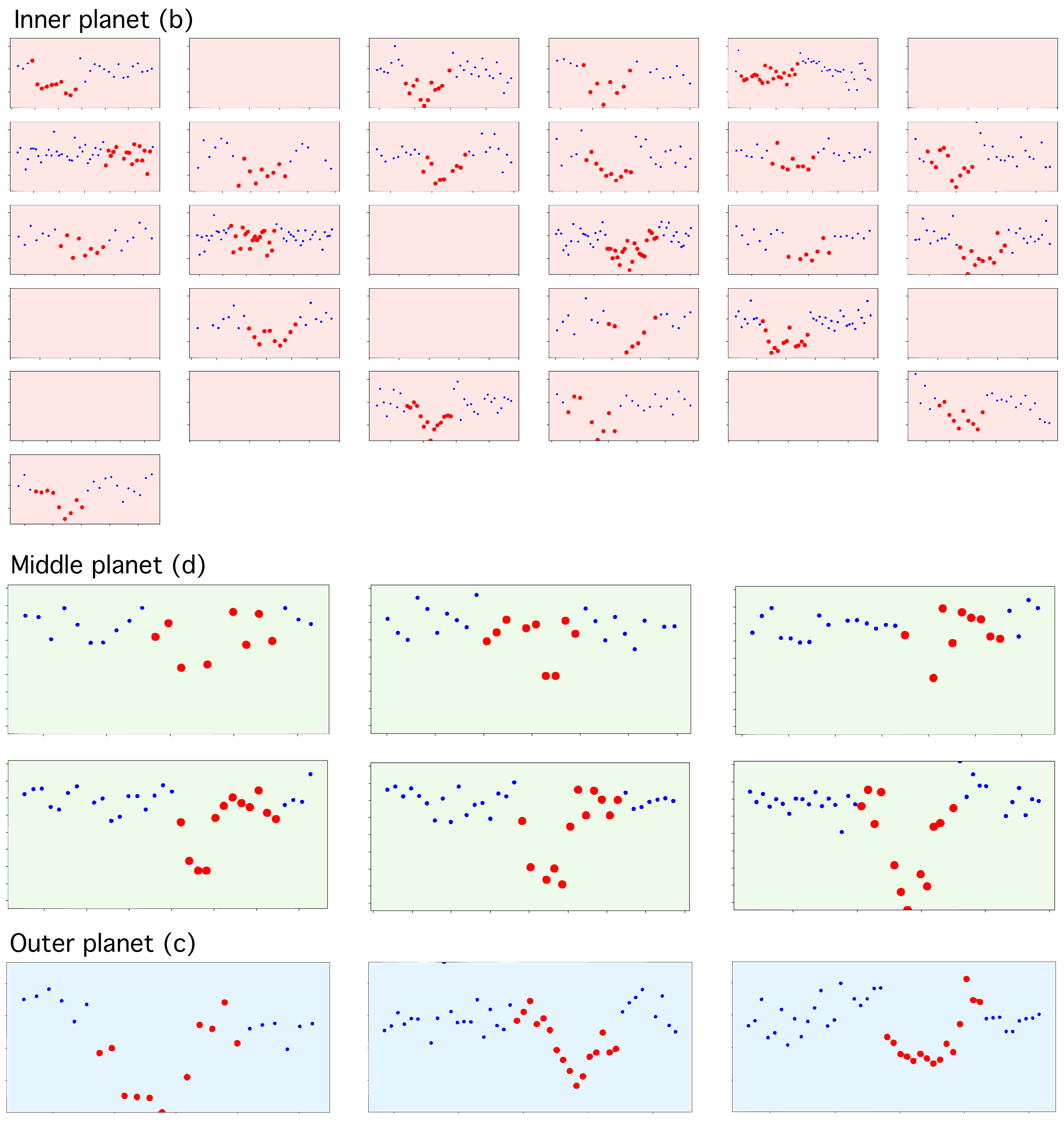}
\caption{An atlas of transits for the inner 49 day planet b (red), the middle 187 day planet d (green) and the outer 303 day planet c (blue). The plots are colour-coded to match the SDE spikes in Fig.~\ref{fig:kepler_47_SDEs}.}  
\label{fig:kepler_47_transit_map}
\end{center}  
\end{figure*} 

 Kepler-47 is the only known transiting multi-planet circumbinary system, containing three \citep{orosz2012b,orosz2019}. From the initial run of \textsc{Stanley} Kepler-47b has the most prominent SDE signal, just above 15. With a radius of $3R_{\oplus}$ it is smallest planet in both the system and indeed of all known circumbinary planets. However, it is the innermost planet with a $49$ day period and so it transits many times more than the outer planets, boosting its detection signal. 

For Kepler-47 we cut out these detected transits of planet `b' from the light curve and then re-run \textsc{Stanley}. A clear detection of the outer-most planet `c' \footnote{Confusingly, the outermost planet is called 'c' and then the middle planet is called `d', since the latter was not confirmed in the original paper \citet{orosz2012b}, but only later by \citealt{orosz2019}.} is then seen, also at an SDE of roughly 15. We repeat the process of removing these detected transits and re-run \textsc{Stanley}, at which point the middle planet `d' is revealed. Curiously it also has an SDE of $\sim 15$. Finally, after removing the transits of `d' there are no additional detectable planets. We show all of the SDE plots in Fig.~\ref{fig:kepler_47_SDEs} and an atlas of the transits of each planet in Fig.~\ref{fig:kepler_47_transit_map}.

In Fig.~\ref{fig:kepler_47_transit_map} it is noticeable that the middle planet has six detected transits which grow deeper over time due to nodal precession, and are only easily visible by eye near the end of the Kepler mission.  When the first paper \citet{orosz2012b} was written only one ``orphan'' transit of this middle planet had been detected, and so it was only a candidate and not confirmed until \citet{orosz2019}. The outer planet transits are more constant in depth. Actually four  transits of `c' exist in the data but one of them is blended with an eclipse, which causes it to get removed by our algorithm. Overall, \textsc{Stanley} is the first algorithm with a demonstrated ability to detect multi-planet circumbinary systems.

\subsubsection{Exceptional case of Kepler-1647}\label{subsubsec:results_known_kepler1647}

 Using the standard criteria of our algorithm, i.e. three transits on the primary star, we do not detect the 1108-day Kepler-1647. The highest peak of the SDE in Fig.~\ref{fig:SDE_plots_WITHtransits} is at 7.5 with a period of 134.8 days. This corresponds to one of the obvious ($0.2\%$) dips in the light curve, folded together with some noise.

To test how {\sc Stanley} performs with looser constraints, we re-run it with a requirement of only two transits. We also change the transit detection to be from the primary star to the secondary star, as the two deep transits for Kepler-1647 are on the secondary, with only one shallow, grazing one occurring on the primary. 

\begin{figure}  
\begin{center}  
\includegraphics[width=0.49\textwidth]{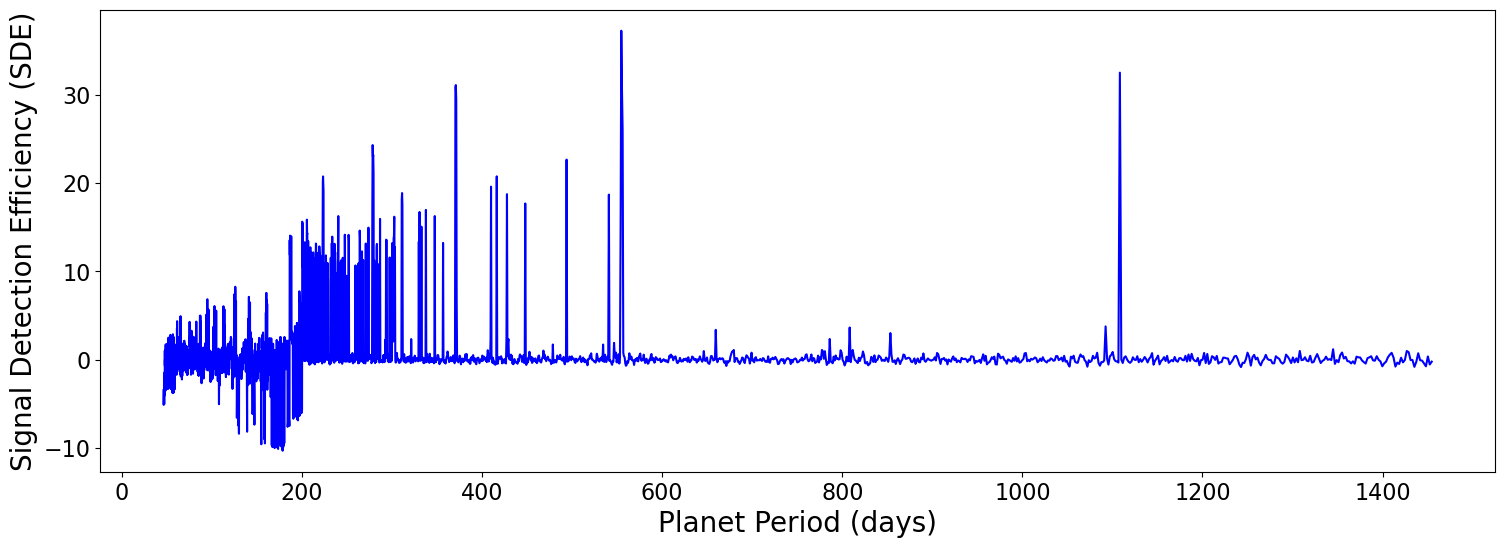}
\caption{SDE for Kepler-1647 with a modified detection criteria of only two transits (instead of three) on the secondary star. The true period of 1108 days corresponds to the second highest peak. The highest peak at roughly half this period is for a model containing the two real transits and a gap transit epoch. We can therefore recover the known planet, but with a degenerate period.}  
\label{fig:Kepler_1647_SDE}
\end{center}  
\end{figure}

The resulting SDE is shown in Fig.~\ref{fig:Kepler_1647_SDE}. \textsc{Stanley} finds the true planet period of 1108 days with a large SDE spike above 30, corresponding to the two deep transits on the secondary star. Curiously, there is a slightly higher SDE peak at roughly half this orbital period. This solution contains the same two transits, plus one transit epoch occurring in a gap. There are also many other ``significant" looking signals that are a result of the weakened two transit criteria.

Ultimately in this case we can say that \textsc{Stanley} has recovered the known planet, but with a degenerate period. The initial discovery of these transits in \citet{armstrong2014} had the same period degeneracy, but the confirmation of the planet in \citet{kostov2016} using a full photodynamical model and the shallow but significant primary transit confirmed the 1108 day period. Given the 554 day model does not contain additional transits, one could argue that Occam's razor favour's the longer-period model, which is why we include it in Table~\ref{tab:known_planets}.

Ultimately, it is up to the user how they use the code and the criteria are programmed to be user-editable.  A two transit criterion will allow for more long-period and marginal planet detections, but at the expense of reliability. Kepler-1647b is also larger than Jupiter and; it is unlikely that small circumbinary planets could be detected with just two transits on typically faint Kepler-targets. Finally, we note that the main aim of the \textsc{Stanley} code is to phase-fold transits with a variable interval, and this is only relevant for three or more transits.

\subsection{Placing detection limits}\label{subsec:results_detection_limits}

\subsubsection{Scaling planet transits}\label{subsubsec:results_detection_limits_scaling}

We artificially reduce the depths of known transits to test \textsc{Stanley's} limits of detectability. To scale the transits we first calculate the  smooth, noise-free trend $\lambda(t)$ of the light curve using a Tukey's biweight filter in \textsc{Wotan}, where the known transits have been masked to be ignored by the detrending filter. The in-transit  lightcurve flux, $f^*(t)$ is then scaled  at each time step according to 

\begin{equation}
\label{eq:scaling_flux}
    f^*(t)= \lambda(t) - (\lambda(t) - f(t))\left(\frac{R_{\rm p}^*}{R_{\rm p}}\right)^2 + \mathcal{N},
\end{equation}
where $f$ is the original flux and $\mathcal{N}$ is Gaussian random noise with a standard deviation equal to

\begin{equation}
    s_{\rm noise} = \left(1-\frac{R_{\rm p}^*}{R_{\rm p}}\right)s_{\rm oot},
\end{equation}
where $s_{\rm oot}$ is the standard deviation of the flux out of transit, after eclipses have been removed and the light curve has been received an initial detrending.

The noise is added to counter-act the effect of the original noise being effectively decreased as the transit is scaled down. In the limiting case of $R_{\rm p}^{*}=0$ the noise added is equal to the noise of the entire light curve.

Note that for planets with visible transits on both primary and secondary stars they both get scaled down proportionally. For the multi-planet system Kepler-47 we are only attempting to detect the innermost planet and hence only scale down these transits, not those of the outer planets.

For each system we scale down $R_{\rm p}$ in steps of $1R_{\oplus}$ and re-run the detection algorithm. Since we are only focused on the known planets, typically near the stability limit, we use a shortened period grid concentrated near the edge of the stability limit: $2.2a_{\rm bin}$ to $4.1a_{\rm bin}$, and a grid in $\theta_{\rm p}$ which is fixed rather than adaptive, which will not have a large effect over a small period range, and if anything will yield a slight under-sampling and hence our results are slightly conservative. This is all for computational expediency since we are not interested in planets up to the original 2 year period limit.

\subsubsection{Finding smaller circumbinary planets}\label{subsubsec:results_detection_limits_finding_smaller}

\begin{figure}  
\begin{center}  
\includegraphics[width=0.49\textwidth]{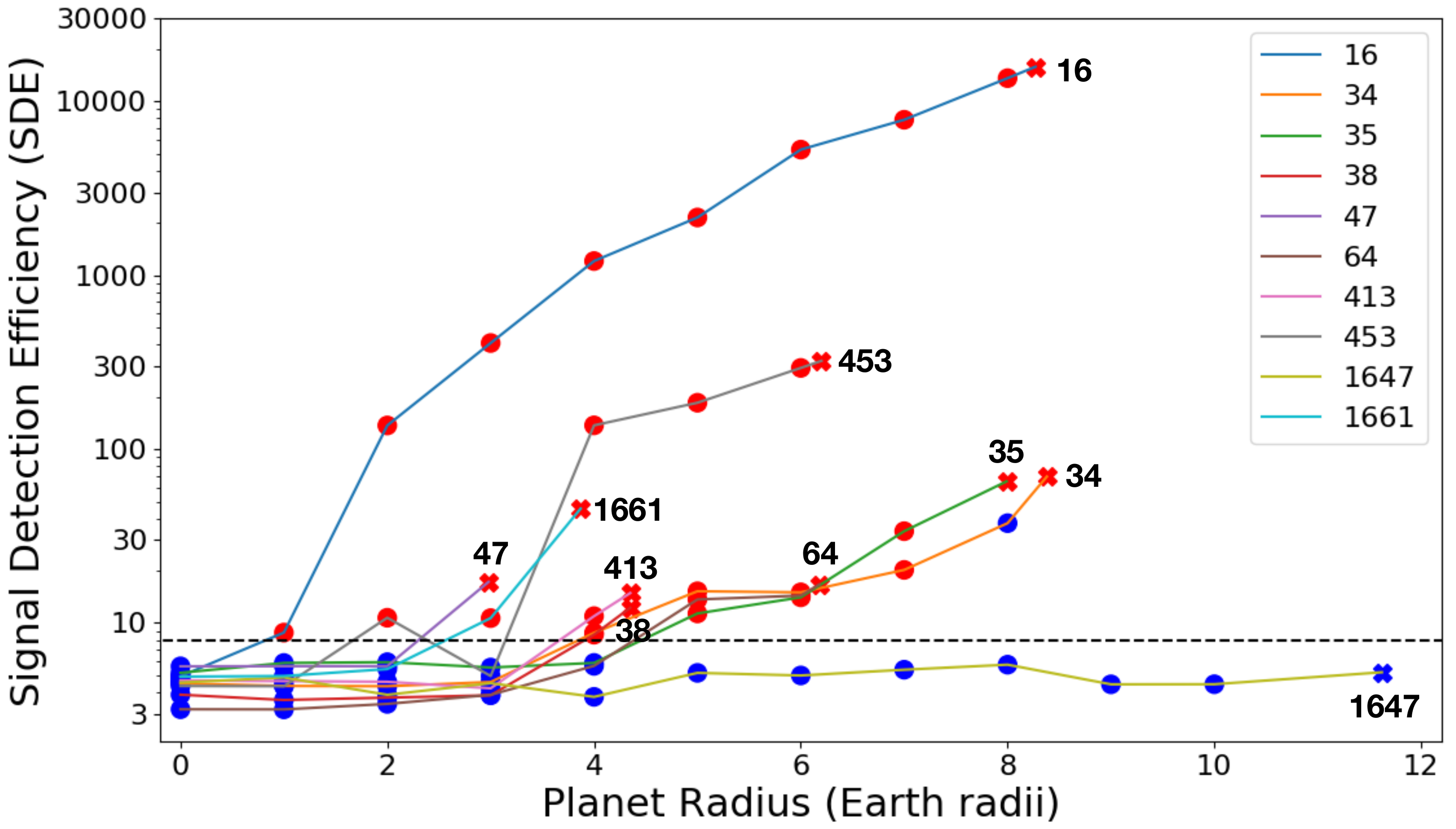}  
\caption{Signal Detection Efficiency (SDE) for known circumbinary planets with reduced transit depths. Each of the ten known circumbinary systems is shown with a different line. The cross is the SDE for detection of the original data (i.e. the value in Table~\ref{tab:known_planets}). For each system the planet radius is reduced from its known value down to $0R_{\oplus}$ in integer steps, which proportionally scales the transit depth according to Eq.~\ref{eq:scaling_flux}. Circles indicate the SDE with scaled transits. Markers are coloured red if more than 50\% (typically more than 80\%) of the primary transits have been recovered by the detection algorithm, and blue otherwise. The black horizontal dashed line at SDE = 8 is a rough limit, above which a detection is considered significant. All of the circles below this line are blue, and all but one above this line are red (see text in Sect.~\ref{subsec:results_detection_limits} for outlier explanation). The SDE is therefore seen to be an accurate indicator of when the algorithm successfully detects the planet transits. Kepler-16 would have been detectable down to $1R_{\oplus}$.}
\label{fig:scaled_SDE}
\end{center}  
\end{figure} 

The detrending and search algorithms are then run normally on these scaled transits. In Fig.~\ref{fig:scaled_SDE} we show the signal detection efficiency (SDE) as a function of the scaled planetary radius. We consider the planet detected if $> 50\%$ of transits are detected, and this is indicated with a red circle (blue otherwise). Each system follows a similar curve, where the SDE decreases as the planet radius is decreased, until it gets below SDE $\sim8$ and the transits are no longer detected above the noise. At $0R_{\oplus}$, i.e. the transits are completely replaced by noise, no detections are made, as should be the case.

Two small exceptions to this are seen. First, for Kepler-34 the transits of the $8R_{\oplus}$ scaled planet are anomalously not detected, whereas they are for smaller radii. This is because Kepler-34 is a roughly equal mass binary and the transits are of similar depth on both stars. For the $8R_{\oplus}$ light curve \textsc{Stanley} has in fact discovered transits of the planet, but curiously four of the five are on the secondary star, and hence it technically gets counted as a non-detection despite a large SDE. Accounting for both primary and secondary transits will be a future improvement (Sect.~\ref{subsubsec:discussion_improvements_secondary_transits}).

The other exception is that for Kepler-453 the $2R_{\oplus}$ and $4R_{\oplus}$ scaled transits are detected but not the $3R_{\oplus}$. That is because this system only has three transits and the detrending algorithm happens to detrend away one of the three transits in the case of $3R_{\oplus}$. It is forever a challenge to develop detrending techniques which remove unwanted signals yet preserve transits. In Sect.~\ref{subsubsec:discussion_improvements_detrending} we discuss a few pieces of future work which could aid this.

Overall we see that our modified SDE is a good predictor of when planets are discovered, and a limit of $\sim 8$ is a reasonable threshold, similar to previous studies on single stars \citep{hippke2019a}.

In Table~\ref{tab:detection_limits} we catalogue the detection limits for each planetary system. We define the smallest detectable planet radius as the one corresponding to an SDE of 8, using a linear interpolation in Fig.~\ref{fig:scaled_SDE}. In most cases we could find planets where the transits have been reduced in depth by over 50\%, and over 90\% in the best cases (Kepler-16 and -453). In four out of nine cases our algorithm pushes into the mini-neptune and super-Earth regime with $R_{\rm p} < 2.5R_{\oplus}$, which are the most common planets around single stars (Fig.~\ref{fig:single_star_abundance}). On the other hand, Kepler-38 was arguably the most difficult planet (could only detect 11\% smaller planets) due to the difficulties arising from the planet's awkward phasing and near 5.5 period commensurability with the binary, as discussed in Sect.~\ref{subsec:results_known}.

\subsubsection{Detection limits as a function of period}\label{subsubsec:results_detection_limits_function_period}

The detection limits derived in Sect.~\ref{subsubsec:results_detection_limits_finding_smaller} are defined for the orbital parameters of the known planet. To derive detection limits at other putative planet periods we take the empirically derived detection limit from Sect.~\ref{subsubsec:results_detection_limits_finding_smaller} and scale it according to

\begin{equation}
\label{eq:detection_limit_period}
    R_{\rm p,lim} \propto P_{\rm p}^{\frac{1}{4}},
\end{equation}
which is derived by assuming the number of data points in transit is proportional to $1/P_{\rm P}$. This is a rough criterion, since we neglect the transit duration being a function of period. This would be simple to account for in single stars, but complicated in binaries where transit durations vary with binary phase (Sect.~\ref{subsec:detrending_transit_duration_theory}), and this affects  both the detrending robustness  and the transit durations and hence SNR of the transit signal. Overall, the detection threshold of a circumbinary system is not sharp; it is highly binary-system dependent, as was seen with Kepler-38 being surprisingly such a challenge.

\begin{figure}  
\begin{center}  
\includegraphics[width=0.49\textwidth]{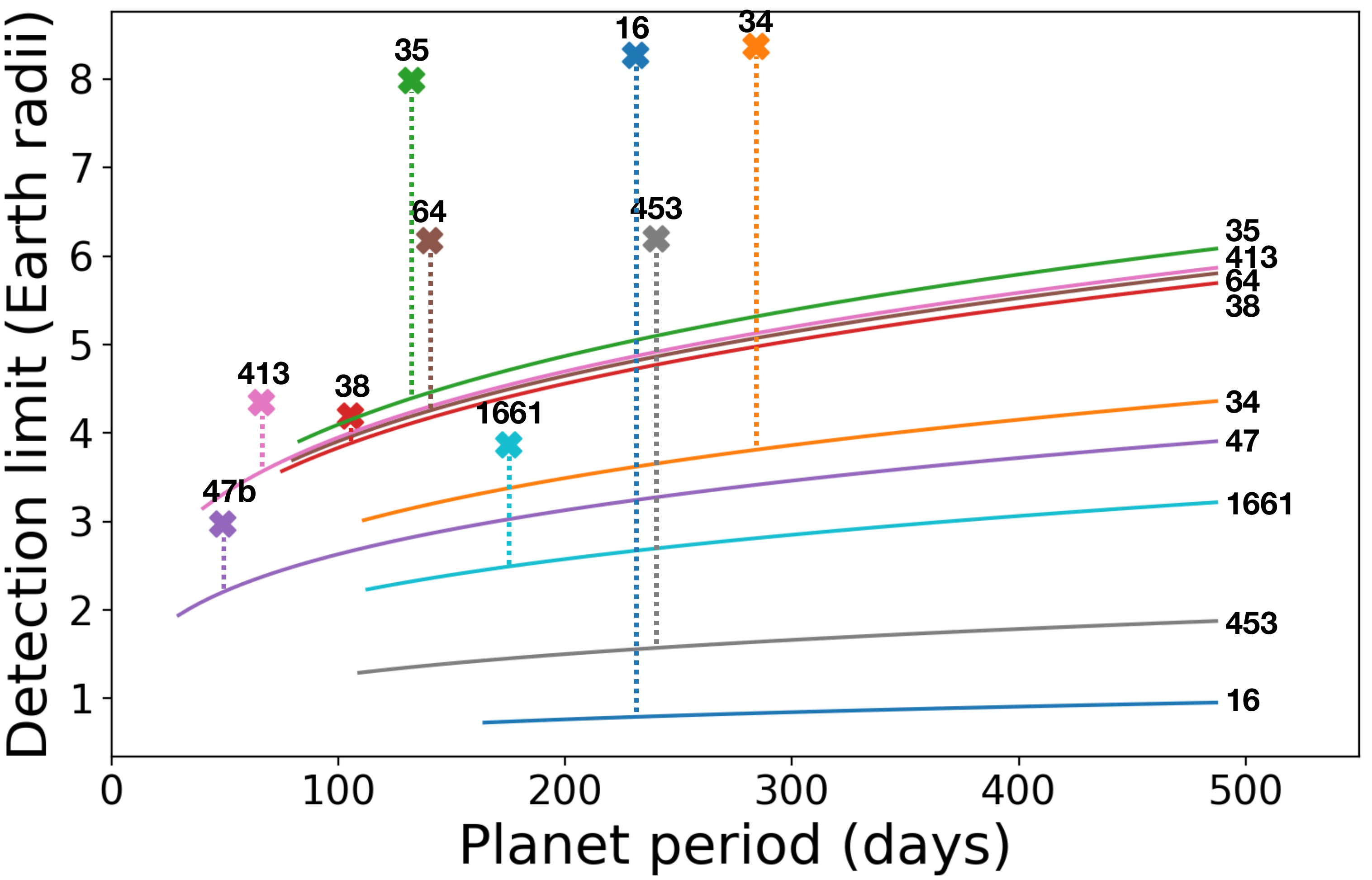}  
\caption{Detection limits for known circumbinary planets as a function of planet period. We derive these curves using the emperically-derived empirical detection limit at the known planetary period (given in Table~\ref{tab:detection_limits}) and scaling it according to Eq.~\ref{eq:detection_limit_period}. Crosses indicate the true radius and period of each known planet, and are connected to the detection limit curves by dotted lines.}
\label{fig:detection_limits_vs_period}
\end{center}  
\end{figure}

In Fig.~\ref{fig:detection_limits_vs_period} we calculate Eq.~\ref{eq:detection_limit_period} over a period range between $4P_{\rm bin}$ (an approximate stability limit based on \citet{holman1999}) and $4/3$ years (guaranteeing at least three transits within the Kepler mission, neglecting gaps in the light curve). The resulting curves, with a shallow dependence on $P_{\rm p}$, are similar to those seen in \citet{deeg2000}. Aside for Kepler-38, -47 and -413, \textsc{Stanley} is sensitive to smaller planets at all stable periods up to 4/3 years.

\begin{table*}[]
\begin{center}
\label{tab:detection_limits}
\begin{tabular}{ccccc}
\hline
Name        & True  &  Detection & Radius       & Transit Depth \\
            & Radius ($R_{\oplus}$) & Limit ($R_{\oplus}$)    & Reduction \% & Reduction \%  \\
\hline
Kepler-16b   & 8.27         & 0.79            & 90           & 99            \\
Kepler-34   & 8.38         & 3.81            & 54           & 79            \\
Kepler-35   & 7.99         & 4.39            & 45           & 70            \\
Kepler-38   & 4.20         & 3.88            & 11           & 20            \\
Kepler-47b  & 3.05         & 2.20            & 28           & 48            \\
Kepler-64   & 6.10         & 4.29            & 31           & 52            \\
Kepler-413  & 4.35         & 3.56            & 18           & 33            \\
Kepler-453  & 6.30         & 1.57            & 75           & 94            \\
Kepler-1661 & 3.87         & 2.49            & 36           & 59 \\
\hline
\end{tabular}
\caption{Detection limits for known circumbinary systems at the known planetary period  based on the recovery of scaled transits. The true radius is taken from the respective discovery paper, and the detection limit is the smallest radius our \textsc{Stanley} algorithm can detect based on scaling the transits to a shallower depth. For Kepler-47b the detection limit is only calculated for the smallest, inner-most planet (see Fig.~\ref{fig:Kepler47Scaling}), but all three planets in this system are detected at an SDE$>15$ (Fig.~\ref{fig:kepler_47_SDEs}). Values are only given for the innermost planet in Kepler-47, and are not produced for the 1108 Kepler-1647, because \textsc{Stanley} does not find with the standard detection criteria (but see Sect.~\ref{subsubsec:results_known_kepler1647}).}
\label{tab:detection_limits}
\end{center}
\end{table*}

\begin{figure*}  
\begin{center}  
\includegraphics[width=0.90\textwidth]{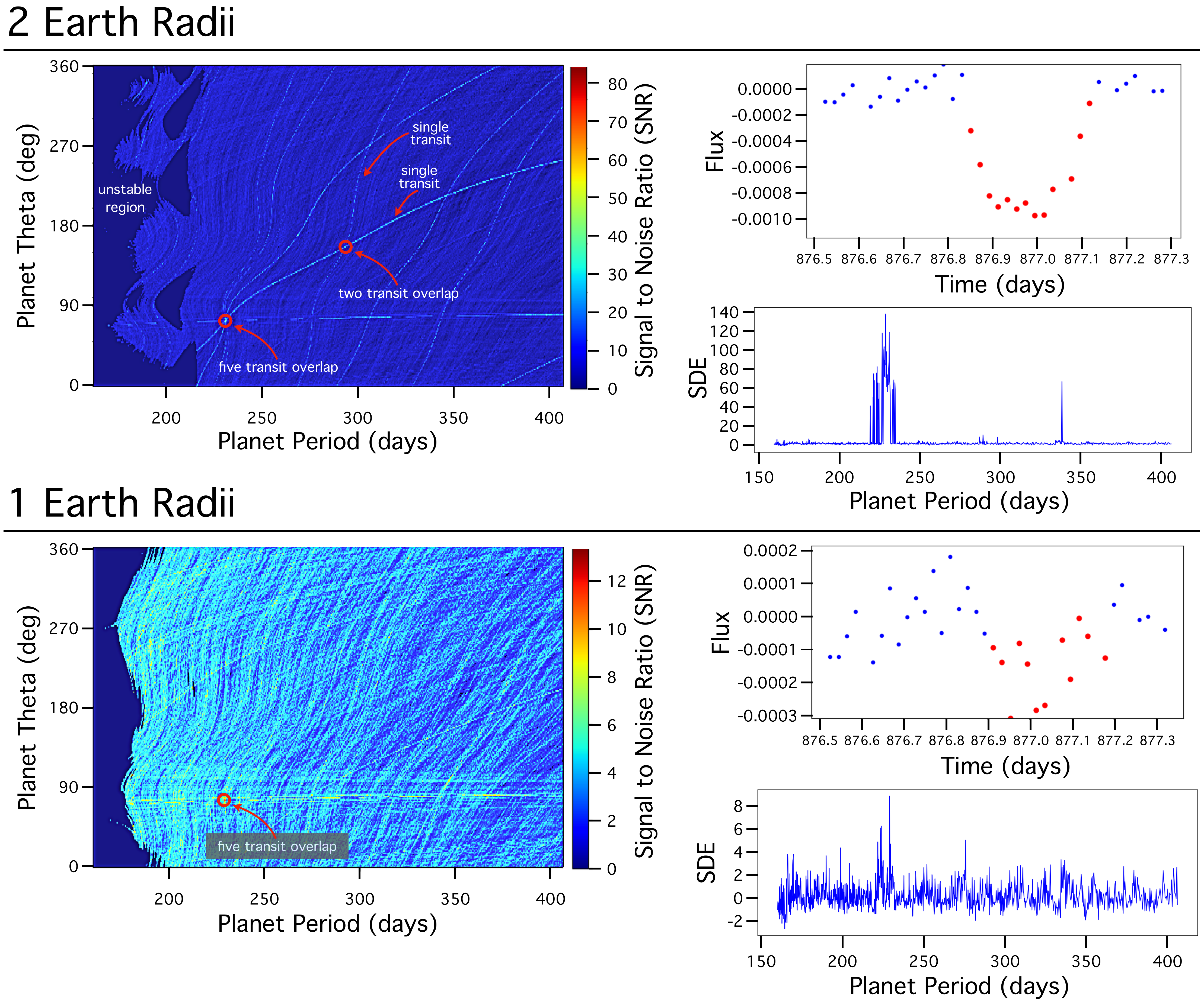}  
\caption{Detection of Kepler-16 with transits scaled according to a $2R_{\oplus}$ (top) and $1R_{\oplus}$ (bottom) planet. The top plots shows the SDE, where the spikes correspond to the same period in both cases but for $1R_{\oplus}$ the SDE of 8.8 is barely above the noise. On the left plots we show the transit fit signal to noise ratio (SNR) from Eq.~\ref{eq:SNR_global_transit} as a function of $P_{\rm p}$ and $\theta_{\rm p}$, calculated for the best-fitting $e_{\rm p}$ and $\omega_{\rm p}$. The plots on the right show an example transit and the SDE. The $2R_{\oplus}$ planet is easily detectable. The $1R_{\oplus}$ planet is detectable just above the SDE threshold of 8. For the $2R_{\oplus}$ planet we annotate various features of the SNR parameter space. On the left the dark blue region corresponds to zero SNR since there are no acceptable solutions here, largely due to instability from being too close to the binary. The light streaks of high SNR correspond to fits including just one of the transits. These streaks are visible because each transit is individually significant. We use red circles to indicate where two and five transits overlap, i.e. the model transit times fit multiple transits. The five transit overlap corresponds to the peak of the SDE. For the $1R_{\oplus}$ planet, we are able to discover the ensemble of five transits but the individual transits are not discernible above the noise.}
\label{fig:Kepler16_scaledSNR}
\end{center}  
\end{figure*}

From Fig.~\ref{fig:scaled_SDE} the smallest detected hypothetical circumbinary planet is  a $1R_{\oplus}$ scaled version of Kepler-16, with six out of seven transits detected and an SDE of 8.8. In Fig.~\ref{fig:Kepler16_scaledSNR} we show how the $1R_{\oplus}$ detection is barely above the noise limit, and compare it to the relatively easy detection of a $2R_{\oplus}$ planet. We have therefore demonstrated that not only an Earth-sized circumbinary planet, but one on a $\sim220$ day orbit in the habitable zone of its binary, is detectable with the \textsc{Stanley} algorithm. We temper this slightly by noting that Kepler-16 is brighter (visual magnitude of 12) and quieter than most Kepler eclipsing binaries, which is probably to be expected for what was the first circumbinary planet discovery.

\begin{figure*}  
\begin{center}  
\includegraphics[width=0.90\textwidth]{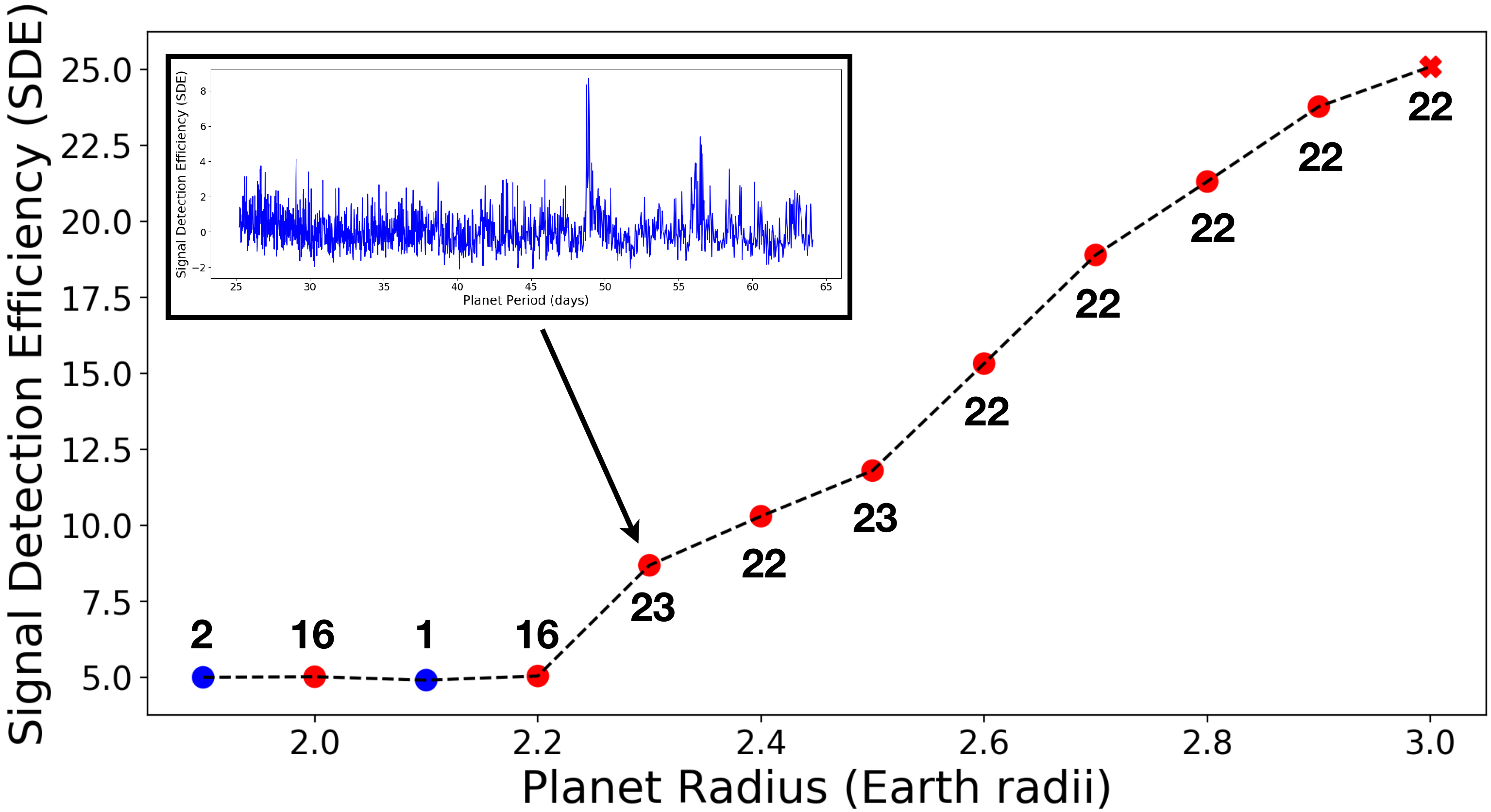}  
\caption{Scaled transits for Kepler-47b, where the deeper, more easily visible transits of the outer two planets have been removed. The number next to each data point indicates how many of the 24 inner transits were detected by \textsc{Stanley}. Like in Fig.~\ref{fig:scaled_SDE}, red markers indicates more than 50\% of transits being found and blue for less. The cross indicates the true $3R_{\oplus}$ radius. The inset figure is the SDE for a $2.3R_{\oplus}$ planet, which is the smallest planet considered detected.}
\label{fig:Kepler47Scaling}
\end{center}  
\end{figure*}

The smallest detected circumbinary planet so far is Kepler-47 at $3R_{\oplus}$. This is strongly detected with SDE = 17.2. However, in the scaled transit simulations in Fig.~\ref{fig:scaled_SDE} no detection was made for $1R_{\oplus}$ or $2R_{\oplus}$. In these simulations we scaled the depths of the inner planet but did not touch the outer planets. We re-run these simulations with the outer planets removed. We consider this reasonable since the outer transits are deeper and easily visible by eye (Fig.~\ref{fig:kepler_47_transit_map}). We also take smaller step sizes of $0.1R_{\oplus}$ to see more precisely what our detection limits are. The results are shown in Fig.~\ref{fig:Kepler47Scaling}.

The smallest detectable  scaled planet around Kepler-47b was $2.3R_{\oplus}$, where 22 out of 24 transits were detected and the SDE = 8.7. The SDE vs $R_{\rm p}$ curve flattens out around an SDE of 5. Curiously, for  two of the solutions considered undetected by the SDE a subset of 16/24 transits were still discovered, but not in a way which distinguishes them from noise, so we do not consider these legitimate detections of the planet.  We also note that Kepler-47 is significantly fainter than Kepler-16 (visual magnitude of 15.4 compared with 12), which would be the main contribution to Kepler-47's weaker detection limit.

\subsection{Searching for additional planets in known systems}\label{subsec:results_additional_planets}

Out of the ten known circumbinary systems, only Kepler-47 has multiple (three) planets \citep{orosz2012,orosz2019}. Around single stars the {\it observed} planet multiplicity is a higher $\sim 20\%$\footnote{Calculated using the Exoplanet Database \url{exoplanetarchive.ipac.caltech.edu/} where as of 19 November 2020 there are 3197 confirmed exoplanet systems, of which 732 are observed to be multiples. Of course the true exoplanet multiplcity rate is likely much higher since observational biases will often cause only a single planet in a system to be observable.}

The inner stability limit requiring $P_{\rm p} \gtrsim 4 P_{\rm bin}$, plus the propensity for planets to orbit binaries  with periods longer than 7 days (a longer period than most Kepler eclipsing binaries) limits where additional planets may stably orbit. However, \citet{quarles2018} show that roughly half of the known systems could host shorter-period planet, squeezed between the known planet and the binary-induced stability limit.  Kepler-1647 in particular is a prime candidate given its 1108 day known planet and 11 day binary. On the other hand, for exterior planets, all but Kepler-1647 could host an additional planet with a period less than two years, which allows us to search companion period ratios at least up to two, which covers most multi-planet systems around single stars \citep{steffen2015}. Under certain conditions co-orbital circumbinary planets (i.e. a 1:1 resonance) are predicted by \citet{penzin2019}, to which the algorithm is also sensitive.

\begin{figure*}  
\begin{center}  
\includegraphics[width=0.99\textwidth]{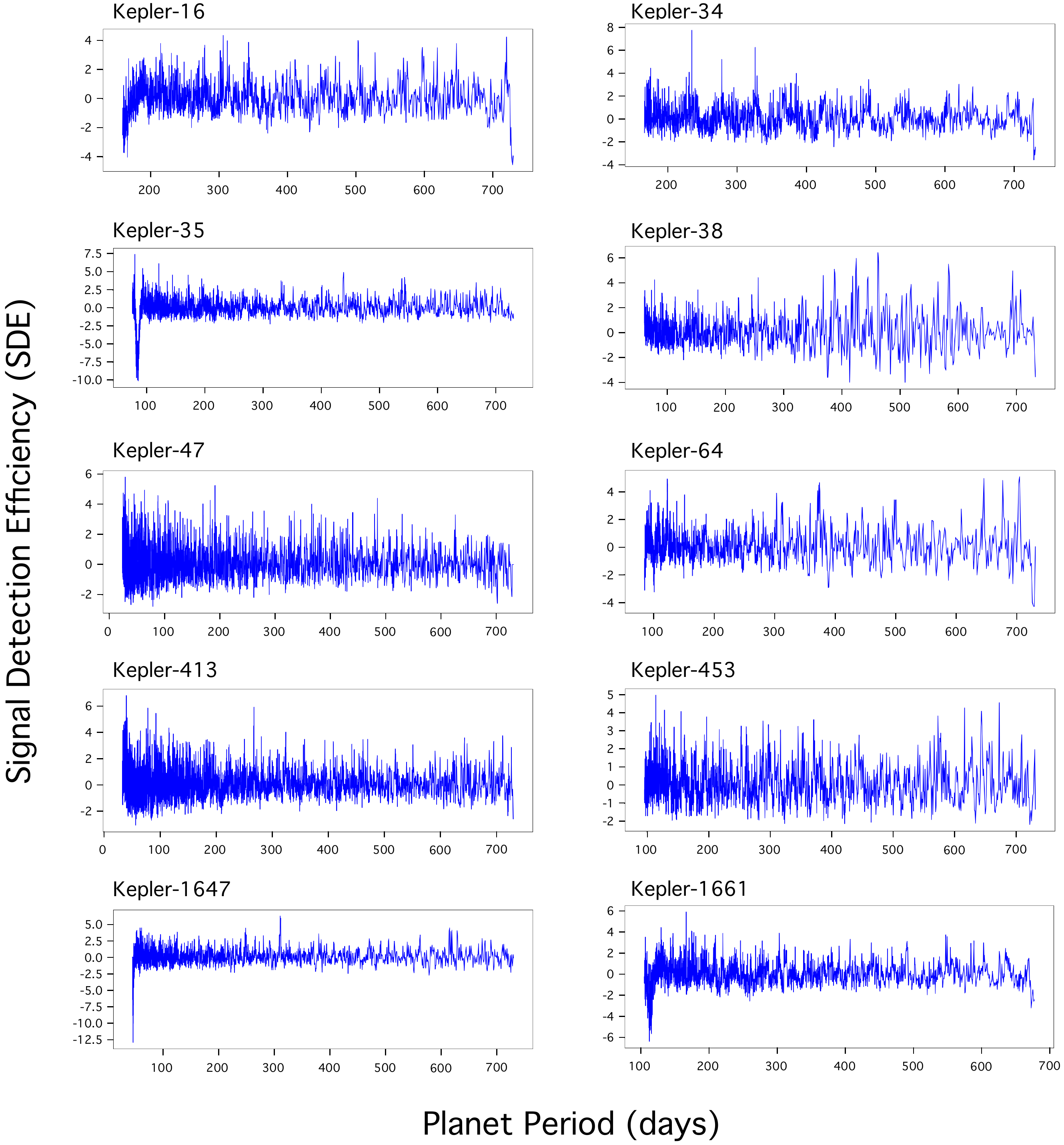}  
\caption{Signal Detection Efficiency (SDE) for all known circumbinary systems with the transits of the known planets removed, in the search for new companions. No significant new detections are made.  The occasional dips in SDE, most prominent in Kepler-35 at $\sim 80$ days and Kepler-1661 at $\sim 120$ days, are created by regions of parameter space where there are few or no  transit models that can be fit to the data, due to stability or failing the criteria of three roughly consistent transits. That produces a large reduction in the SNR in that parameter space compared with surrounding regions, leading to a significantly negative SDE. They are not signs of ''inverse'' or ''upwards'' spikes in the data, such as might be created by flares or microlensing events.}
\label{fig:companion_planet_search}
\end{center}  
\end{figure*}

For all ten systems we cut out all known transits by eye and then re-run the search algorithm. The results, by means of the SDE, are shown in Fig.~\ref{fig:companion_planet_search}. No new planets are detected.

The detection limits as a function of period in Fig.~\ref{fig:detection_limits_vs_period} show our rough sensitivity to interior and exterior additional planets, although we caution that the phasing of transits is also important for their detectability.  In each system in this figure we can rule out $4R_{\oplus}+$ interior planets. For five of the systems this interior limit is better than $3R_{\oplus}+$ and for three of the systems we were sensitive to $2R_{\oplus}+$. For Kepler-16 we could have detected planets slightly smaller than Earth on both interior and exterior orbits out to almost 500 day periods. In every system in Fig.~\ref{fig:detection_limits_vs_period} we can rule out exterior planets larger than $6R_{\oplus}$, and for many of the systems this limit is significantly smaller. We also rule out the commensurate planets predicted by \citet{penzin2019}, unless they are significantly smaller than the detected planet at that period.

It is possible that large companion planets do exist but on misaligned orbits that did not transit during Kepler. However, it is almost guaranteed that any such planets would eventually enter transitability \citep{martin2015a,martin2017}, and indeed this happened with the middle planet in Kepler-47 \citep{orosz2019}. Analysis of the TESS revisit of the Kepler field may reveal such new planets \citep{kostov2020b}.

\section{Discussion}\label{sec:discussion}

\subsection{Comparison with other past methods}\label{subsec:discussion_comparison}

Here we provide a qualitative comparison with some past methods for finding circumbinary planets. The codes for these other methods are typically not published so an in depth quantitative comparison of their efficiency is not possible at this time.

\subsubsection{Traditional box least squares}\label{subsubsec:discussion_comparison_bls}

For completeness we include traditional BLS (Kovacs \& Mazeh) in this list, but the requirement of no or at least minimal transit timing variations makes it inappropriate for circumbinary planets. Without modifications, it could only be expected to find planets with coincidentally regular transit intervals, e.g. from a period commensurability with the binary or for a short period planet exhibiting many transits, some of which randomly happen to be regularly-spaced. The BLS algorithm would be aided by short binary periods, reducing the TTVs, although at this point the planet transits may actually be of longer duration than the binary orbit, adding complications.

\subsubsection{Jenkins 1996}\label{subsubsec:discussion_comparison_jenkins}

Eclipsing binaries were considered as early as \citet{borucki1984} as an ideal target for transit surveys, since the fact that they eclipse geometrically biases the transit probability. \citet{schneider1990,schneider1994} expanded on this idea, including the possibility of misaligned orbits. 

\citet{jenkins1996} and its later implementation in papers such as \citet{deeg1998}, to our knowledge, created the first actual algorithm dedictated to transiting circumbinary planet detection. Despite being the oldest circumbinary-specific algorithm in this list, it is in fact conceptually the most similar to our own. \citet{jenkins1996} uses a ``Matched Filter Method'', which is a cross-correlation between the light curve and a large bank of model transit light curves. It is a brute force approach, using an N-body algorithm to produce the transit models, like  in our paper. Their figures 13 and 14 are very similar to our Fig.~\ref{fig:Kepler16_scaledSNR}. Their search grid of orbital elements is limited to the planet period and starting phase, but they do discuss the implications for neglecting other effects such as eccentricity. Unlike our \textsc{Stanley} algorithm, \citet{jenkins1996} explicitly tries to match the depth of transits by including in their model a variable planet radius. The means of deducing a statistically significant detection also differ from our paper

Their work was specifically targeted at 1.27 day eclipsing binary CM Draconis. It consists of two M-dwarfs, which is a beneficial but rare configuration; small planets are more easily detectable. \citet{jenkins1996} demonstrate a sensitivity down to $1.4R_{\oplus}$. Their grid size consisted of about 40,000 unique transit models, which is a few orders of magnitude smaller than the grids we typically use in this paper\footnote{Although we have the benefit of two decades worth of computing improvements.}. We are not aware of any published application of this algorithm to a broader sample.

\subsubsection{CBP-BLS - Ofir 2008}\label{subsubsec:discussion_comparison_ofir}

The \citet{ofir2008} algorithm - CBP-BLS - applies BLS to a light curve that has been treated to account for the barycentric motion of the binary. This shifts the location of potential circumbinary transits based on the binary phase. Indeed, the barycentric motion of the binary is the largest source of transit timing and duration variations, and by removing it the remaining light curve will be closer to strict periodicity. It does neglect, however, non-Keplerian perturbations which can shift transits by more than a transit duration. These unaccounted for effects will blur any detection. No application of CBP-BLS to a new planet search or the recovery of known planets has been published, although conference proceedings \citep{ofir2015} state that the planets discovered at that time were all recoverable.

\subsubsection{Quasiperiodic Automated Transit Search - QATS - Carter \& Agol 2013}\label{subsubsec:discussion_comparison_qats}

The QATS algorithm by \citet{carter2013} is designed to find transiting planets with large transit timing variations. It has largely been applied to multi-planet systems around single stars, where in special cases mean motion resonances can make the TTVs on the order of hours. In theory, QATS can be applied to almost arbitrarily high TTVs, such as the extreme case of circumbinary planets where the TTVs may be $\sim 5-10\%$ as large as the transit interval \citep{armstrong2013}.

In QATS instead of fitting a fixed transit interval (or equivalently, an orbital period) you fit a maximum and minimum transit interval. At each transit epoch the explicit transit time can vary between these bounds. The downside to this method is that there is no physical basis for the variations; the variations are simply whatever best fits the data. This hampers detection reliability, and makes the method susceptible to detrending errors, particularly for long period planets \citep{orosz2012b}. 

We do note that our transit detection algorithm does use one ``QATS-esque'' element. Specifically, the three-times widened transit window within which we find the lowest flux (Sect.~\ref{subsec:search_transitmask}). However, whilst this sliding over a widened window is purely data driven, like QATS, it is a small correction relative to the overall circumbinary TTVs, which are physically determined by the N-body code.

No large circumbinary planet search with QATS has been published, but the algorithm has made important contribution to planets around single stars (e.g. Kepler-36, see \citealt{agol2018}).

\subsection{Kostov et al. 2013}\label{subsubsec:discussion_comparison_kostov}

 \citet{kostov2013} was the first to present a novel application of the BLS algorithm (Sect.~\ref{subsubsec:discussion_comparison_bls},  \citealt{kovacs2002}) to circumbinary planets. The light curve is split up into segments with length corresponding to roughly a few times the binary orbital period, as this was a reasonable period at which to expect transiting planets based on both the stability limit and the known discoveries at the time of Kepler-16b, -34b, -35b, -38b and -47bc. Then within each segment BLS is run to find the most significant pair of transit-like dips. Given the segment length, this would correspond to consecutive transits on the same star (most likely the primary) separated by roughly the planet's period, and not a transit on each star separated by a few days. This method allows BLS to be applicable since the irregular transit intervals of circumbinary planets are not apparent when BLS is only used to find a single interval.

The \citet{kostov2013} algorithm then repeats this process with the flux of the light curve flipped, such that the BLS algorithm is therefore finding upwards ``anti-transits''. A plot is produced showing the depth of all of the transits and anti-transits in the different segments of the light curve. If there is nothing but white noise then one would expect as many of each type of event, with roughly the same range of depths. If there are significant transits then these will stand out as being of greater depth. If the number of outlier dips in the light curve is greater than some merit function\footnote{Based on a method implemented by \citet{burke2006} in the search for planets around single stars.} then the algorithm triggers a human eye search to confirm that the transit signal is real and coherent across the entire light curve and not just pairs of transits in short segments. \citet{kostov2013} portrays this as a ``semi-automated'' algorithm, although realistically most transiting candidates undergo a human inspection to some degree.

This algorithm was applied to yield the first discovery of Kepler-64, which was independently discovered by Planet Hunters using solely by-eye methods \citep{schwamb2013}. \citet{kostov2013} also presented and independent discovery and characterisation of the then recently published Kepler-47b and c. Impressively, the algorithm was also able to detect a $1.5R_{\oplus}$ transiting super-Earth injected into the Kepler-16 light curve, with 75\% of the transits recovered. Improvements were later made to the method to make it applicable to misaligned planets with gaps in the transit sequence. This led to the discovery of Kepler-413 in \citet{kostov2014}.

\subsubsection{Armstrong et al. 2014}\label{subsubsec:discussion_comparison_armstrong}

\citet{armstrong2014} is best known as the first comprehensive calculation of the  occurrence rate of circumbinary planets, using the Kepler mission. However, it also contains a novel method for detecting circumbinary planets, and indeed found three circumbinary planet candidates which were later confirmed: Kepler-453 \citep{welsh2015}, Kepler-1647 \citep{kostov2016}, and Kepler-1661 \citep{socia2020}. The method is to phase-fold the light curve on a fixed period and then search for significant dips within a wide window. Some transits would coincidentally be coherently stacked, but most would only be bunched together. The window was not arbitrarily chosen, but based on the TTV model derived in the (\citealt{armstrong2013}, with inspiration from \citealt{agol2005}). These TTVs account for both the binary's barycentric motion and apsidal precession of the planet.

\citet{armstrong2014} is one of only two large systematic circumbinary planet searches using an automated algorithm (the other being \citealt{klagyivik2017}, Sect.~\ref{subsubsec:discussion_comparison_klagyivik}). The consistency of the method allows for reliable  occurrence rate constraints. The method is not, however, particularly sensitive to small planets. Indeed, constraints could only be made on planets down to $4R_{\oplus}$ (the $3R_{\oplus}$ Kepler-47b could not be recovered), and admittedly with large error bars near this limit.

\subsubsection{Klagyivik et al. 2017}\label{subsubsec:discussion_comparison_klagyivik}

\citet{klagyivik2017} is a similar undertaking to \citet{armstrong2014}; a new method was both developed and applied to a large data set, in this case CoRoT, and constraints were placed on the  occurrence rate of circumbinary planets. Their method may be considered a hybrid between QATS \citep{carter2013} and \citet{armstrong2014}. A transit model is fitted allowing for variable intervals. This variation is fitted as a free parameter, but unlike in QATS for which the TTV bounds are arbitrary, \citet{klagyivik2017} fit TTVs within the physical limits defined by the circumbinary TTV geometry (derived in \citealt{armstrong2013}). \citet{klagyivik2017} can therefore coherently stack the transits, unlike \citet{armstrong2014} who bunched them in within a wide window.

No planets were discovered from the CoRoT data and detection limits could only be placed down to $4R_{\oplus}$. In addition to the poorer photometry than the Kepler mission, the short baseline of CoRoT is challenging for circumbinary detection, given their tendency for relatively long period orbits \citep{munoz2015,martin2015,hamers2016,fleming2018}. Overall, it would be interesting to see this method applied to Kepler.

\subsection{QATS EB - Windemuth et al. 2019b}\label{subsubsec:discussion_comparison_windemuth}

The most recently published method by \citet{windemuth2019b} combines and expands several past techniques. First, the geometric positioning of the binary and planet at each transit epoch is accounted for using two independent Keplerians, creating a ``regularized'' light curve. The planetary orbit is assumed to be circular. The second step is to apply QATS, which accounts for any additional TTVs caused either physically (e.g. eccentric planets and non-Keplerian variations such as precession) or due to imperfect orbital and stellar parameters in the model.
Compared with vanilla QATS, the transit timing models have at least some physical basis and are not completely arbitrary. 

Compared with \textsc{Stanley}, the physical basis for the TTVs in QATS EB (independent Keplerians) is not as robust as using an N-body  algorithm, but it does make the QATS EB algorithm significantly faster to run. \citet{windemuth2019b} quote $\sim 5$ minute run times per target on a single CPU, compared with ours taking typically tens or hundreds of hours. In fact, \citet{windemuth2019b} presents their algorithm as a ``compromise between the two extremes'' of vanilla QATS and a fully brute force grid search such as ours.

\citet{windemuth2019b} recover known gas giants Kepler-35 and Kepler-64, with what they define as a SNR of 26.1 and 35.6, respectively. This is essentially the same as our SDE definition, which Table~\ref{tab:known_planets} lists as 65.2 and 16.7 for these two planets, respectively. However, our SDE peaks, as noted in Sect.~\ref{subsec:search_sde}, get broadened by searching over a grid of $e_{\rm p}$ and $\omega_{\rm p}$, whereas Windemuth assume circular planets. A closer comparison is is to calculate our SDE for the best-fitting $e_{\rm p}$ and $\omega_{\rm p}$, which yields SDEs of 68.6 and 26.7 for Kepler-35 and Kepler-64, respectively. Our Kepler-35 detection is therefore more significant whereas our Kepler-64 detection is a bit less. We note that differences may not be a function of the detection algorithm but rather different detrending methods.

\citet{windemuth2019} also demonstrate a SNR = 20 detection of an injected $1R_{\oplus}$ planet in a system with orbital parameters similar to Kepler-47b. However, there are two reasons why this is not comparable to our results in Sect.~\ref{subsubsec:results_detection_limits_finding_smaller}, for which $2.3R_{\oplus}$ was the smallest detectable planet. First, rather than using the true Kepler-47 light curve and scaling the transits, a synthetic light curve is created with white noise and injected transits. Given that a large part of the challenge in detecting small circumbinary planets is filtering out both physical and instrumental variations, the assumption of white noise is unrealistic. Second, the white noise used has a standard deviation of $5\times10^{-5}$, which is $\sim 8$ times smaller than the actual photon noise on the Kepler-47 light curve (calculated as the standard deviation of the out of eclipse and transit flux after detrending). So whilst QATS-EB is demonstrated to have a sensitivity to Earth-sized planets, it is in a highly idealized setting. Work is presently being undertaken to apply the  \citet{windemuth2019b} algorithm to a larger sample of Kepler eclipsing binaries (Windemuth, priv. comm.).

\subsubsection{The human eye}

It would be remiss of us to neglect the one method that  has actually yielded all of the transiting circumbinary planets: actually has yielded new circumbinary planet publications - the human eye. Led in particular by Jerry Orosz, Bill Welsh, Veselin Kostov and Laurence Doyle, all of the known planets were painstakingly found by visually examining thousands of light curves. There has also been success with Planet Hunters, which crowd sources many human eyes. One of their first discoveries was the circumbinary planet Kepler-64b (\citealt{schwamb2013}, and independently also published by \citealt{kostov2013}). The complex transit signature of circumbinary planets is well suited to this platform.

We believe that our new method has advantages, but it is yet to be seen if our method can challenge the human eye for new discoveries.

\subsection{Possible improvements to \textsc{Stanley}}\label{subsec:discussion_improvements}

\subsubsection{Deeper optimization of the detrending efficiency}\label{subsubsec:discussion_improvements_detrending} 

To detrend our light curves we combine existing methods with bespoke procedures specific to circumbinary planets. Some of our methods are admittedly ad hoc, and whilst they ultimately result in an algorithm that achieves its goal of small circumbinary planet detection, our methods could no doubt be more rigorously optimized. We used the biweight and cosine filters from \textsc{Wotan} since they are amongst the best for single stars, but it might be worth testing different filters with different window lengths. More specifically, one could run the detection algorithm from start to finish with modified detrending and quantify planet detection efficiently, similar to the tests run in \textsc{Wotan} \citep{hippke2019b}. The multi-dimensional nature of both the eclipsing binary detrending and the planet detection would make this computationally challenging though.
    
\subsubsection{Transits on both primary and secondary stars}\label{subsubsec:discussion_improvements_secondary_transits}

Our algorithm as it stands only detects transits on the primary star. In all but one of the circumbinary systems the planet exhibits multiple transits on the primary star and those  alone would be enough for confirmation\footnote{The exception is Kepler-1647b, which has one primary and two secondary transits, but its 1108 day  period would likely be too long for a small circumbinary planet search anyway.} In fact, most of the  known circumbinary systems have secondary transits that are either invisibly shallow or geometrically non-existent \citep{martin2019a}. Accounting for both transits could improve the SNR for near equal mass binaries such as Kepler-34, but for most binaries it would likely be folding noise onto the signal. Given the current method of timing transits by iterating the N-body integrator back and forward, timing secondary transits would also slow down the detection algorithm.  A future improvement to this code will be to at least give the user the option of folding both primary and secondary transits, preferably without a significant slow down to the algorithm. The choice of whether or not to include secondary transits may be informed by whether or not the secondary eclipse depths are comparable to that of the primary.

\subsection{Future applications}\label{subsec:future_work}

\subsubsection{Systematic survey for small circumbinary planets in Kepler using the Windemuth catalog} 

Paper II in preparation is an application of this new \textsc{Stanley} algorithm to the over 700 eclipsing binaries with stellar masses and radii derived in the Windemuth catalog \citep{windemuth2019}. It is both a search for new planets and a new, tighter constraint on the circumbinary planet  occurrence rate.

\subsubsection{Application to the entire Kepler eclipsing binary catalog}

The entire Villanova EB catalog totals almost 3000 targets.  Whilst some of these will likely remain inappropriate for planet searches, such as heartbeat stars, ellipsoidal variables and very tight contact binaries, we hope to expand our sample beyond the initial $\sim700$ Windemuth targets. Those not in the Windemuth catalog typically do not have accurately derived masses and radii though. As demonstrated in this paper though, circumbinary transits can be found even if the stellar parameters are a little bit ($\sim 10-20\%$) wrong. Furthermore, the algorithm can scan over a grid of stellar parameters too. There remains hope then that the algorithm can be applied to find circumbinary transits even when the binaries are poorly constrained.

\subsubsection{Application to TESS}

The short, typically one month observing windows of TESS are challenging for the discovery of circumbinary planets, which to date have been found on orbits of 50 days and typically much longer. Some TESS targets closer to the ecliptic pole receive longer windows, but none to the same extent as Kepler's four years. To discover planets with longer periods than the observational window we may exploit the concept of a ``1-2 punch'', where a planet on a single conjunction transits both primary and secondary stars, which can be exploited to better characterise the planet than one of the same period around a single star \citep{kostov2020b}. Adapting this algorithm to discovery planets with less transits, using an instrument with more variability and a smaller aperture, is a challenging future task.

\subsubsection{Public release}

The \textsc{Stanley} code is in the process of being finalised for public release. This involves making the code cleaner and well-commented, adding in depth documentation and testing it on a variety of systems. Please contact the authors for early access.

\section{Conclusion}\label{sec:conclusion}

We have presented \textsc{Stanley}: a new automated algorithm to detrend eclipsing binaries and then find circumbinary planets. By phase-folding on variable transit times and durations, which are physically modelled with an N-body  code, the algorithm coherently stacks transits. This allows the detection of small planets with shallow transits which may be individually indistinguishable above the noise.

We significantly recover all known circumbinary systems,  including all three planets in the Kepler-47 system and the 1108 day Kepler-1647b (but only if the detection criteria are loosened). For each system we scale down the transit depths and can find significantly smaller planets than those found by eye, with the best being Kepler-16, detectable to a sub-Earth radius. We show also search for new companion planets but find none, which means that they either do not exist or are significantly smaller than the known gas giants.

This paper is a precursor to Paper II, where we apply this algorithm to over 700 Kepler eclipsing binaries to hunt for new circumbinary planets. We will also derive improved  occurrence rate constraints on the circumbinary planet population, which can then be compared with that around single stars.

\section{Acknowledgements}\label{sec:acknowledgements}

We thank the many people over the years with whom conversations have motivated and shaped this research, including Dave Armstrong, Scott Gaudi, Greg Gilbert, Vedad Hodzic, Veselin Kostov, Rosemary Mardling, Ben Montet, William Petersen, Amaury Triaud, Jerry Orosz, Stanley Trowbridge, St\'ephane Udry and Bill Welsh, and indeed the entire exoplanet groups at Geneva, Chicago and The Ohio State.  In addition, we thank  Trevor David, Michael Hippke and Ren\'e Heller for comments on a draft of this paper, in particular the methods of detrending. We also appreciate Diana Windemuth and Eric Agol for open discussions about their similar efforts to find small circumbinary planets. Thank you to Nora Bailey for helping get \textsc{Stanley} running on the U Chicago cluster, and apologies for hogging it so much. This research would not have been possible without several packages developed by the community:  \textsc{astropy} \citep{astropy2013}; \textsc{astroquery} \citep{ginsberg2019};  \textsc{celerite} \citep{foreman-mackey2017}; \textsc{Lightkurve} \citep{lightcurve2018}; {\sc Rebound} \citep{rein2012,rein2015}; \textsc{Wotan} \citep{hippke2019b}. DVM has been funded by two fellowships from the Swiss National Science Foundation (P2GEP2\_171992 and P400P2\_186735). 

\bibliography{sample63.bib}{}
\bibliographystyle{aasjournal}



\end{document}